\documentclass{SciPost}

\usepackage{physics}
\usepackage{graphicx}    
\usepackage[normalem]{ulem}
\usepackage{dcolumn}     
\usepackage{bm}          
\usepackage{mathbbol}    
\usepackage{mathtools}
\usepackage{dutchcal}    
\usepackage[dvipsnames]{xcolor}
\usepackage{xkcdcolors}
\usepackage{hyperref} 
    \hypersetup
    {   
        unicode=true,           
        pdftoolbar=true,        
        pdfmenubar=true,        
        pdffitwindow=false,     
        pdfstartview={FitH},    
        pdfauthor={Gabriele Bellomia, Carlos Mejuto-Zaera, Massimo Capone, Adriano Amaricci},
        pdftitle={Symmetry-resolved quasilocal entanglement as a key to
Mott and pseudogap physics in plaquette CDMFT},
        pdfkeywords={strong correlations, entanglement measures, quantum information theory, dynamical mean-field theory, quantum embedding, cluster methods, open quantum system, metal-insulator transition, pseudogap metal},
        pdfnewwindow=true,      
        colorlinks,
        linkcolor={red!50!black},
        citecolor={blue!50!black},
        urlcolor={blue!80!black}
    }

\renewcommand\labelenumi{(\roman{enumi})}
\renewcommand\theenumi\labelenumi

\newcommand\eg{\mbox{\it e.g.\ }}
\newcommand\ie{\mbox{\it i.e.\ }}

\long\def \gb#1{{\color{black}{#1}}}

\usepackage[bitstream-charter]{mathdesign}
\DeclareSymbolFont{usualmathcal}{OMS}{cmsy}{m}{n}
\DeclareSymbolFontAlphabet{\mathcal}{usualmathcal}

\fancypagestyle{SPstyle}{
\fancyhf{}
\lhead{\colorbox{scipostblue}{\bf \color{white} ~SciPost Physics }}
\rhead{{\bf \color{scipostdeepblue} ~Submission }}

\fancyfoot[C]{\textbf{\thepage}}
}

\begin{document}

\pagestyle{SPstyle}

\begin{center}{\LARGE \textbf{\color{scipostdeepblue}{
Symmetry-resolved quasilocal entanglement as a key to
Mott and pseudogap physics in \emph{plaquette} CDMFT
}}}\end{center}

\begin{center}\textbf{
{\small
Gabriele Bellomia\textsuperscript{1,2,$\star$},
Carlos Mejuto-Zaera\textsuperscript{3}, 
Massimo Capone\textsuperscript{1,4} and
Adriano Amaricci\textsuperscript{4}
}
}\end{center}

\begin{center}
{\small
{\bf 1} Scuola Internazionale Superiore di Studi Avanzati, {SISSA}, Trieste, Italy
\\
{\bf 2} Institute of Solid State Physics, TU Wien, Wieden, Vienna, Austria
\\
{\bf 3} Laboratoire de Physique Th\'eorique, CNRS, Universit\'e de Toulouse, Toulouse, France
\\
{\bf 4} Istituto Officina dei Materiali, Consiglio Nazionale delle Ricerche, CNR--IOM, Trieste, Italy
\\[\baselineskip]
$\star$ \href{mailto:email1}{\small gabriele.bellomia@tuwien.ac.at}
}
\end{center}

\section*{\color{scipostdeepblue}{Abstract}}
\textbf{\boldmath{%
In order to identify intrinsic quantum resources in strongly interacting fermionic 
systems, we introduce a symmetry resolution of the parity-superselected
two-site entanglement, as measured by 
both the fermionic negativity and the relative entropy of entanglement. 
We discuss how the symmetry resolution can be extended also to the full
entanglement in pure and mixed two-site states.
We apply our recipe to the Hubbard model, first for a dimer, and then for a plaquette cluster dynamical mean-field theory, an established method tailored to accurately capture short-range correlations. 
Following the evolution of the model through the interaction- and density-driven Mott transitions at zero temperature, we highlight how the three symmetry-resolved components of the two-site entanglement probe different key aspects of Mott and pseudogap physics.
}}

\vspace{\baselineskip}



\vspace{10pt}
\noindent\rule{\textwidth}{1pt}
\tableofcontents
\noindent\rule{\textwidth}{1pt}
\vspace{10pt}
\setcounter{secnumdepth}{3}
\setcounter{tocdepth}{3}


\section{Introduction}

In the last decades, entanglement properties of many-body systems
have become established both as central tools for the study of phases of matter
and quantum critical points 
\cite{Srednicki_arealaw,EE_1stpaper,Osterloh_Nature,EntanglementRMP,Calabrese_EEreview,Laflorencie_Review,piflux_maksim},
and as fundamental resources for quantum information processing and computation
\cite{HHHH_RMP,QResources_RMP,QuantumSupremacy,QuantumAdvantage}.

Systems of interacting electrons are expected to provide special resources associated with electron-electron correlations, but they entail additional subtleties,
as a clear
and universal definition of entanglement between indistinguishable particles 
is still lacking \cite{Benatti2020,Szalay2021}.
In the framework of fermionic mode entanglement \cite{Banhuls2007,Friis2013,Friis2016}, namely entanglement 
between \emph{distinguishable} modes, particles enter only via the occupation 
numbers of these modes. Hence, it is significantly nontrivial to separate the 
contributions to mode entanglement that are genuinely provided by the electron-electron 
repulsion from those that may be present also in noninteracting systems. 
For instance, a highly delocalized noninteracting metal features a rich entanglement
structure in real-space, which can be clearly traced back to simple hopping processes.
These hopping-induced entanglement properties can be useful to characterize 
phase transitions of non-interacting fermions, but hardly describe genuine quantum resources, as the system is
trivially diagonalized via a one-body rotation -- \eg a Fourier transform for translational invariant systems. In the rotated basis there is no mode entanglement, hence no quantum resource.

Some progress in this respect has been achieved by employing parity and charge
superselection rules (SSR) 
\cite{Wick1952,Wick1970,Bartlett2003,Cirac2004a,Cirac2004b,Dowling2006,SSR_lecturenotes},
which -- at least for small subsystems -- effectively remove the hopping amplitudes 
from the density matrix and isolate the portion of real-space entanglement that is 
independent of charge delocalization.
Furthermore, the SSR-constrained portion of mode entanglement has been proven to be
physically and operationally accessible to quantum information protocols, ensuring its
usefulness as a quantum resource in both quantum chemistry
\cite{DissotiationParadoxSchilling,EntanglementVsCorrelation2021,Schilling_QuantumScience,ding2022_faithful_entanglement,ding2023physical_entanglement,Faraday_Natural_Orbitals} 
and lattice models for solid-state systems
\cite{MonkmanSirker2020,Galler_SSR,DelMaestro_SSR_spinless,DelMaestro_SSR_attractive,Bellomia2023,BellomiaPhD,Canio_orbital_concurrence,thermal_death_ssr,SSR_PG_ColdAtoms}.

In this work, we propose a further resolution of the SSR entanglement of lattice fermions into physically distinct terms which helps to identify the quantum resources associated with distinctive phases of correlated fermionic systems. 
In particular, we consider the single-band Hubbard model, a model that combines a fundamental character with its relevance for different transition
metal oxides, including copper- and nickel-based high-$T_\text{c}$ superconductors
\cite{Arovas_HubbardReview,Gull_HubbardReview,Kitatani2020_Nickelates}, making it an ideal playground to assess the ability of our symmetry-resolved entanglement to capture the intrinsic quantum resources of phases such as Mott insulators and pseudogap states.
The Hamiltonian reads
\begin{equation}
H = -t\sum_{\langle{ij}\rangle\sigma} 
    \biggl[ c^\dagger_{i\sigma}c_{j\sigma} + \text{h.c.}\biggr] 
    + U\sum_i \left(n_{i\uparrow}-\frac{1}{2}\right)\left(n_{i\downarrow}-\frac{1}{2}\right) - \mu N,
\label{eq:hubbard}    
\end{equation}
where $ c^\dagger_{i\sigma}$ ($c_{i\sigma}$) create (annihilate) an electron 
with spin $\sigma=\uparrow,\downarrow$ at site $i$. $t$ is the uniform and 
isotropic hopping amplitude between nearest neighbor (NN) sites 
$\langle{ij}\rangle$, $U$ is the local interaction strength and 
$n_{i\sigma}= c^\dagger_{i\sigma} c_{i\sigma}$ is the local occupation per spin.
At zero temperature, the chemical potential $\mu$ fixes the total conserved 
charge of the system $N=\sum_{i\sigma} n_{i\sigma}$. This conservation law 
ensures a precise block structure in all reduced density matrices 
\cite{bellomia_local_classical} and their partial transpose 
\cite{Imbalance_negativity_1st_paper,Murciano_imbalance_negativity_free_fermions,Murciano_imbalance_negativity_nonequilibrium}. From the latter it is possible to define a concept of
symmetry-resolved entanglement, in which the total bipartite entanglement
between an A and a B pair of subsystems is decomposed in contributions 
labeled by the \emph{charge imbalance} $\Delta N = N_\mathrm{A}-N_\mathrm{B}$.
As for other notions of symmetry-resolved entanglement \cite{GoldsteinSela2018,Xavier2018_equipartition,Capizzi_sym-resolved_QRE_CFT},
it has been shown that, under fairly general conditions, each charge-imbalance
sector contributes equally to the full entanglement. While this property is 
remarkably useful to estimate efficiently the entanglement of large subsystems,
it does not help unveil possibly distinct sources of entanglement, that
characterize at a fundamental level the different, competing phases of 
strongly-correlated systems.

For the special case of two-site Hubbard subsystems, intensely studied in the 
last few years \cite{DissotiationParadoxSchilling,ding2023physical_entanglement,Bellomia2023,Roosz_2RDMfrom2GF,Bippus_Two-Site_Negativity_Hubbard,SSR_PG_ColdAtoms},
we develop an alternative symmetry resolution, with contributions labeled
by a \emph{local charge jump}, instead of the charge imbalance. 
The 0-charge jump entanglement arises from antiferromagnetic fluctuations
in the two-site density matrix. It essentially quantifies the efficiency
of purification in Bell pairs of the spin-singlet kind
$\frac{1}{\sqrt{2}} \Bigl(\mid\uparrow\rangle \otimes \mid\downarrow\rangle -
\mid\downarrow\rangle \otimes \mid\uparrow\rangle \Bigr)$.
The other SSR-allowed contribution is the 2-charge jump entanglement,
which corresponds to Bell pairs of entangled holons ($\ket{\bullet}$)
and doublons ($\ket{\uparrow\downarrow}$), in the form
$\frac{1}{\sqrt{2}} \Bigl(\ket{\bullet} \otimes \mid\uparrow\downarrow\rangle +
\mid\uparrow\downarrow\rangle \otimes \ket{\bullet} \Bigr)$.
For these two classes of entanglement the symmetry resolution is strictly 
rigorous as it directly descends from the block structure of the superselected 
two-site density matrix. On the other hand, the hopping-induced entanglement, 
which is generated by all the off-diagonal matrix elements that are forbidden
by superselection rules, can be unambiguously associated to a 1-charge jump 
sector only for pure two-site states. For mixed two-site states, we define a
hopping entanglement by exploiting the (pseudo)additive properties of two 
considered entanglement measures, the fermionic negativity
\cite{Peres96,Horodecki97,Vidal_Negativity_def,Plenio_Negativity_log,FermionNegativity,FermionicNegativity_monotonicity,ThermalNegativity,TwistedNegativity} 
and the relative entropy of entanglement \cite{Vedral_RMP}.

\smallskip

Using cellular dynamical mean-field theory (CDMFT)
\cite{DMFT_RMP,CDMFT_ORIGINAL,Kotliar2001,Biroli2002}, we study zero-temperature solutions of 
the two-dimensional Hubbard model without symmetry breaking.
We show that the weakly-correlated metals
found at either small interaction strength or large doping fractions are 
largely dominated by hopping-induced entanglement and lack any finite 
spin-singlet entanglement. The latter does instead dominate the quasilocal
(confined to nearest neighbors) entanglement of the paramagnetic Mott 
insulator at half filling and the pseudogap metal at small doping fractions.
Finally, we relate the holon-doublon entanglement to the holon-doublon 
mechanism, previously proposed to drive the interaction-driven and
density-driven Mott transitions
\cite{Watanabe2006_holon_doublon_binding,Zhou2014_holon_doublon_binding,Prelov2015_holon_doublon_binding,Capello_VMC_Mott}.

\smallskip

This paper is organized as follows.
In Section \ref{sec:entanglement} we briefly review the two entanglement
measures we will employ throughout the work, as well as the central
underlying object: the fermionic partial transpose of the two-site density
matrix. In Section \ref{sec:sym-res} we present in detail our symmetry
resolution of the two-site entanglement, for both pure and mixed states, and
provide some exact results to compare the main properties of the 
antiferromagnetic, holon-doublon and hopping-induced entanglement measures,
in a well-controlled setting.

Readers interested in the results for the
two-dimensional Hubbard model may go directly to 
Section \ref{sec:cdmft_results}. In this case, we suggest
to keep in mind that the antiferromagnetic and holon-doublon entanglement are
associated to the two Bell pairs written above, but do not simply quantify the
projection of a given CDMFT solution into those quasilocal states. In general,
the mixing of spin singlets and spin triplets will dampen or even cancel out 
the antiferromagnetic entanglement, while the mixing of the holon-doublon pair
with fully empty and fully occupied two-site states will suppress the holon-doublon
entanglement. The hopping-induced entanglement arises from a much more complicated
interplay between many different off-diagonal amplitudes, but can always be understood
in terms of single electron hopping processes, as the name suggests. 
For checking the technical details behind the precise definition of the symmetry 
resolution, the reader can selectively refer to the appropriate
subsections of Section \ref{sec:sym-res}.
A brief discussion of the relationship between our symmetry resolution and the
more traditional charge-imbalance resolution of the fermionic negativity can be
found in Appendix \ref{app:imbalance}.

Finally, Appendices \ref{app:cdmft_ed} and \ref{app:risb} discuss technical
details of our implementation of CDMFT with an exact diagonalization (ED) solver
and its limitations in the weak-coupling regime -- with respect to alternative 
cluster embedding theories.

\section{Quantifying entanglement between two Hubbard sites}
\label{sec:entanglement}

In this paper, we introduce entanglement measures of pure 
and mixed states of two Hubbard sites, either isolated or
part of a larger system. Due to spin-degeneracy,
the local Hilbert space of each Hubbard site has dimension
2. Hence, the two-site Hilbert space has dimension 4 and
the associated Fock space has dimension 16.
These simple 4-qubit states have been previously
studied in the context of molecules 
\cite{DissotiationParadoxSchilling,EntanglementVsCorrelation2021,Schilling_QuantumScience,Faraday_Natural_Orbitals} 
and also in fermionic lattice
models for condensed matter, as the most accessible
description of spatial (two-point) entanglement
\cite{ding2022_faithful_entanglement,ding2023physical_entanglement,Bellomia2023,BellomiaPhD,Roosz_2RDMfrom2GF,Bippus_Two-Site_Negativity_Hubbard}.

A two-site density matrix describes two uncorrelated 
single-site states if it can be expressed as a product
state $\rho_{ij} = \rho_i \otimes \rho_j$, where $i$ and $j$
label the two given sites. Product states are also the 
only possible pure unentangled states.
This means that for pure two-site states the entanglement 
between the two sites fully determines all spatial correlations
\cite{Vedral2001,Discord&Classical_unified,TenNet_RMP}.

\begin{figure}
    \centering
    \includegraphics[width=0.42\linewidth]{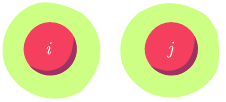} \\
    $E_{ij}=0 \enspace\Longleftrightarrow\enspace \rho_{ij} = \sum_k \lambda_k \rho_i^k \otimes \rho_j^k$ \\[1ex]
    \includegraphics[width=0.42\linewidth]{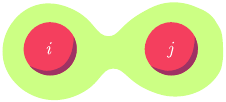} \\
    $E_{ij}>0 \enspace\Longleftrightarrow\enspace \rho_{ij} \neq \sum_k \lambda_k \rho_i^k \otimes \rho_j^k$  
    \caption{Visualization of a mixed two-site
    state, described by a Hubbard Hamiltonian. The associated
    density matrix $\rho_{ij}$ is assumed to be mixed because of
    the embedding in a larger lattice and/or thermal effects.
    It will be pure only in the case of an isolated dimer, at
    zero temperature.
    The two-site entanglement $E_{ij}$, defined in
    terms of a bipartition in single sites $i$ and $j$,
    is associated to the quantum
    spatial correlations of the state, as
    sketched by the green shades: Disconnected for
    a separable two-site state (top), in which the two sites
    are only classically correlated. Connected 
    for an entangled one (bottom), which is described by a 
    single quantum state.}
    \label{fig:sketch}
\end{figure}

In mixed two-site states, however, it is possible to have purely
classical correlations, such as those introduced by thermal
effects or the embedding into a lattice. In general, distinguishing
classical correlations from genuine entanglement, which grants
essential resources for quantum information processing and
computation \cite{HHHH_RMP,QResources_RMP,QuantumSupremacy,QuantumAdvantage}, 
is a hard computational problem \cite{Separable_Set,REE_NPcomplete}. 
One of the most quantitative approaches relies on determining
the distance between the given mixed density matrix and the
set of mixed \emph{separable} (unentangled) states
\cite{Horodecki2001,Plenio_EntanglementMeasures,EntanglementDetection_review}. 
These are defined as classical mixtures of product states, so that 
the separable set is the \emph{convex hull} of the uncorrelated set.
For a two-site system, it reads
\begin{equation}
    \mathrm{Sep}(i,j) = \left\{\,\rho_{ij} \,\,\Bigr|\,\,\, \rho_{ij} = \sum_k \lambda_k \rho_i^k \otimes \rho_j^k, \enspace \lambda_k \geq 0, \enspace \sum_k \lambda_k = 1 \,\right\}.
    \label{eq:sep}
\end{equation}
It follows naturally that entangled two-site states are defined as
those not in $\mathrm{Sep}(i,j)$. We define the two-site
entanglement $E_{ij}$ as the distance between the given
$\rho_{ij}$ and the closest state in $\mathrm{Sep}(i,j)$
\cite{Plenio_EntanglementMeasures}. The concrete quantification
of $E_{ij}$ obviously depends on the quantum metric
in the space of two-site density matrices.
In Fig.\,\ref{fig:sketch} we sketch the fundamental 
distinction between separable and entangled two-site states,
corresponding to $E_{ij}=0$ and $E_{ij}>0$, regardless of
residual classical correlations.

\smallskip

Remarkably,
in the special case of a pure two-site density matrix -- 
relevant for an isolated dimer at zero temperature -- 
the problem of quantifying entanglement is beautifully 
solved by the bipartite von Neumann entropy 
\cite{Srednicki_arealaw,EE_1stpaper,Osterloh_Nature,EntanglementRMP,Calabrese_EEreview,Laflorencie_Review},
which determines
exactly the $E_{ij}$ distance, under any reasonable choice of
quantum metric. Hence, for a pure dimer, we can evaluate
the two-site entanglement as $E_{ij} = S_i = S_j$, where $S_i$
and $S_j$ denote the von Neumann entropy of the single-site 
reduced density matrices $\rho_i = \Tr_j \rho_{ij}$ and 
$\rho_j = \Tr_i \rho_{ij}$ (sometimes called local entropy
\cite{Franca_LDAentropy,XiDai2013,Cocchi_ColdAtoms,WalshLetter2019,WalshPRXQuantum2020,Franca_local_entropy_superc}).
We will explicitly compute this
quantity for the half-filled ground state of the Hubbard dimer
in Section \ref{sec:dimer}. But for a generic mixed state, the
local entropy is not a measure of entanglement \cite{EntanglementDetection_review,Plenio_EntanglementMeasures}.

\subsection{Positive partial transpose and fermionic negativity}
\label{sec:negativity}

For any choice of quantum metric in the space of all two-site density matrices, computing
$E_{ij}$ is a hard, open problem, unless $\rho_{ij}$ is pure
\cite{ding2022_faithful_entanglement,ding2023physical_entanglement,EntanglementDetection_review,Plenio_EntanglementMeasures}.
A popular and to some extent effective alternative is given by the entanglement negativity \cite{Vidal_Negativity_def,Plenio_Negativity_log},
directly associated to the positive partial transpose (PPT) criterion \cite{Peres96,Horodecki97}.

We recall that any density matrix has a positive semidefinite (nonnegative) spectrum,
and that the standard matrix transpose is a spectrum preserving operation. 
Thus, we can readily show that any
separable two-site state keeps its nonnegative spectrum under a \emph{partial} transpose,
namely a transposition of the degrees of freedom associated to only one of the two sites. 
Explicitly
\begin{equation}
    \rho_{ij} \geq 0 \enspace\land\enspace \rho_{ij} \in \,\mathrm{Sep}(i,j)
    \enspace\Longrightarrow\enspace
    \rho_{i(j)^\intercal} \equiv 
    \sum_k \lambda_k \rho_i^k \otimes (\rho_j^k)^\intercal \geq 0,
\end{equation}
as any $(\rho_j^k)^\intercal$ must have the same nonnegative spectrum as
$\rho_j^k$, and $\rho_i^k$ is unaffected by the partial transpose operation.
This proves the direct PPT separability criterion \cite{Peres96},
valid for both pure and mixed separable states as a \emph{necessary} condition
for separability. Namely, the presence of at least one negative
eigenvalue in $\rho_{i(j)^\intercal}$ always detects an entangled state.
The correspondent sufficient condition, or reverse PPT criterion,
such that a positive partial transpose $\rho_{A(B)^\intercal}\geq0$ 
implies that $\rho_{AB}$ is separable has been proven
only for small-enough $A$ and $B$ subsystems \cite{Horodecki97}. 
Specifically, the requirement is that $\dim{\rho_A} \leq 2$ and $\dim{\rho_B}\leq3$
(or, of course, the analogue condition with $A$ and $B$ swapped).
In fact, many mixed states are known, 
for larger Fock spaces, that violate the reverse PPT criterion. In this case the
entanglement is considered to be ``hidden'' (or bound) \cite{Bound_Entanglement}.
We highlight that for the fermionic two-site density matrices which we are interested in,
defined in a 16-dimensional Fock space, the presence of bound entanglement cannot be excluded. 
In Sections \ref{sec:sym-res-pure} and \ref{sec:sym-res-mixed} we will show how our 
proposed symmetry resolution of the partial transpose $\rho_{i(j)^\intercal}$ allows 
to apply the PPT criterion to some smaller diagonal blocks, for which the PPT 
criterion is both necessary and sufficient.
\smallskip

The PPT criterion, in itself, provides only a binary statement:
a given density matrix has either positive partial transpose or not. 
Yet, an entanglement monotone can be defined
from the partial transpose 
\cite{Vidal_Negativity_def,Plenio_Negativity_log}, fulfilling many
properties of a quantitative entanglement measure 
\cite{EntanglementDetection_review,Plenio_EntanglementMeasures,Horodecki2001}
(besides the possible hidden entanglement in high-dimensional Fock spaces). 
The so-called entanglement negativity of a density matrix $\rho_{ij}$ is defined as
\begin{equation}
    \mathcal{N}_{ij} = 
    \log(1 + \sum_k \sigma_k(\rho_{i(j)^\intercal}) - \Tr\rho_{i(j)^\intercal}),
    \label{eq:negativity_svd}
\end{equation}
where $\sigma_k(\rho)$ indicates the singular values of $\rho$, obtained from its singular value decomposition (SVD).
For a normalized density matrix $\Tr\rho_{i(j)^\intercal}=1$, and Eq.\,\ref{eq:negativity_svd} reduces to the logarithm
of the trace norm of the partial transpose; the explicit trace term keeps the definition
meaningful also for the non-normalized diagonal blocks of $\rho_{i(j)^\intercal}$.
For bosonic systems, the partial transpose is Hermitian,
so that the sum over the SVD in 
Eq.\,\ref{eq:negativity_svd} can be reformulated as a sum over the
negative eigenvalues of $\rho_{i(j)^\intercal}$, namely
\begin{equation}
    \mathcal{N}_{ij} = 
    \log(1 + \sum_k \biggl[ \left|\lambda_k(\rho_{i(j)^\intercal})\right| - 
    \lambda_k(\rho_{i(j)^\intercal}) \biggr]), \label{eq:negativity_eig}
\end{equation}
where $\lambda_k(\rho)$ indicates the eigenvalues of $\rho$.
This is the origin of the name for $\mathcal{N}_{ij}$. 
For fermionic systems, besides the obvious
signs needed to put the basis states in canonical ordering for the partial
transposition, additional phases have been proposed 
\cite{FermionNegativity,FermionicNegativity_monotonicity,ThermalNegativity,TwistedNegativity}
to be applied on the transposed elements. These phases ensure that, as for
the case of bosons, the partial transpose corresponds to a time reversal in
the Wigner function formulation of quantum mechanics \cite{FermionNegativity}.
There are indeed known cases in which the standard partial transpose fails to
detect entanglement that is instead revealed by applying these additional phase
factors \cite{FermionicNegativity_monotonicity}. The correspondence between
partial transposes and time reversal operations has been recently proposed to
give a clear physical interpretation of how the negativity measures entanglement:
by partial transposing the density matrix, spatial correlations which violate 
Bell inequalities are mapped into causal correlations which cannot be replicated by 
spacelike separated systems \cite{PPT_as_space-time_swap}. In practice, off-diagonal
elements that can introduce spatial entanglement are transformed to quantum amplitudes
describing temporal fluctuations on each subsystem.

The formal definition of a fermionic partial transpose is not unique, as the special
additional phases can be chosen in different, nonequivalent, ways \cite{TwistedNegativity}.
Yet, the fermionic negativity, computed as a sum of singular values, is indeed unique,
and can be proven to satisfy the monotonicity conditions for entanglement measures as
for the standard partial transpose of bosonic systems \cite{FermionicNegativity_monotonicity}.
Among the different choices, we define the fermionic partial transpose for a two-site
subsystem as
\begin{equation}
\begin{aligned}
    \rho_{i(j)^\intercal}^\mathrm{F} &= \displaystyle\sum_{\Psi_i^\lambda}\sum_{\Psi_i^\nu}\sum_{\Psi_j^\lambda}\sum_{\Psi_j^\nu}
    \langle{\Psi_i^\lambda\Psi^\lambda_j}|{\rho}|{\Psi_i^\nu\Psi^\nu_j} \rangle
    |{\Psi_i^\lambda}\rangle\langle{\Psi_i^\nu}| \otimes 
    \Bigl(|{\Psi_j^\lambda}\rangle\langle{\Psi_j^\nu}|\Bigr)^{\color{red!50!black}\!\intercal} \\
    &=\displaystyle\sum_{\Psi_i^\lambda}\sum_{\Psi_i^\nu}\sum_{\Psi_j^\lambda}\sum_{\Psi_j^\nu}
    \langle{\Psi_i^\lambda\Psi_j^{\color{red!50!black}\nu}}|{\rho}|{\Psi_i^\nu\Psi_j^{\color{red!50!black}\lambda}}\rangle
    |{\Psi_i^\lambda}\rangle\langle{\Psi_i^\nu}| \otimes |{\Psi_j^\lambda}\rangle\langle{\Psi_j^\nu}|
    {\color{red!50!black}\exp\!\left(i\pi\phi^{\lambda\nu}_{ij}\right)},
\end{aligned}
\label{eq:fermionic_ppt}
\end{equation}
where
\begin{equation}
    \phi^{\lambda\nu}_{ij} = \tfrac{1}{2}{\left[\left(\mathrm{P}(\Psi_j^\lambda)+\mathrm{P}(\Psi_j^\nu)\right)\bmod 2\right]} + {\left(\mathrm{P}(\Psi_i^\lambda)+\mathrm{P}(\Psi_i^\nu)\right)}\times{\left(\mathrm{P}(\Psi_j^\lambda)+\mathrm{P}(\Psi_j^\nu)\right)}, \label{eq:fermionic_ppt_phase}
\end{equation}
and $\mathrm{P}\bigl(|n_1, n_2, \dots \rangle\bigr) = \Bigl(\displaystyle\sum_i n_i \Bigr)\!\!\mod 2\,$ is the parity of the given Fock state $|n_1, n_2, \dots \rangle$.

Finally, we define the two-site fermionic negativity as
\begin{equation}
    \mathcal{N}_{ij}^\mathrm{F} = 
    \log(1 + \sum_k \sigma_k\left(\rho^\mathrm{F}_{i(j)^\intercal}\right) - \Tr\rho^\mathrm{F}_{i(j)^\intercal}).
    \label{eq:fermionic_negativity}
\end{equation}

\subsection{Relative entropy of entanglement (REE)}
\label{sec:qre}

There are many different choices for a metric in the space
of quantum states \cite{Multipartite_Geometric,GeometricEntanglementPRA,GeometricQuantumInfo_intro,GeometricQuantumInfo_book,GeometricQuantumInfo_review}.
Among them, the quantum relative entropy (QRE) has emerged as one of the most effective \cite{Vedral97,Vedral98,Vedral2000,Vedral_RMP}. 
The QRE between a state $\rho$ and a state $\rho'$ is defined as
\begin{equation}
    S(\rho||\,\rho') = \Trace \left[
        \rho \left(\log\rho - \log\rho'\right)\right].
\end{equation}
Despite not being symmetric in its arguments and not fulfilling 
the triangle inequality (hence being only a \emph{pseudo}-metric), 
the QRE can be minimized analytically with 
respect to the set of uncorrelated (product) states.
The resulting minimum corresponds to the
von Neumann \emph{mutual information} \cite{Vedral97}, 
which has a clear operational meaning as it bounds
from above all possible correlation functions \cite{Wolf}.
Moreover, one can prove that the minimum QRE from the set of 
classically correlated states -- albeit not obtainable
in closed form -- corresponds exactly to the \emph{quantum}
\emph{discord} 
\cite{Discord_def,Discord_REE,Discord_NPcomplete}, 
an established hallmark of the quantum character of bipartite
correlations \cite{Vedral2001,Discord&Classical_unified}.

Similarly, one can define the \emph{relative entropy of entanglement}
(REE) as the minimal QRE between a given state and the set of 
bipartite separable states (see Eq.\,\ref{eq:sep}). Finding the 
closest separable state under the QRE metric is a formidable open
problem for a general mixed state \cite{REE_NPcomplete}, and exact
expressions for the REE exist only in a few special cases. 
For instance, the REE of a pure state corresponds to the von Neumann
entropy of either reduced state, and some closed expressions have
been derived for specific bipartite systems in presence of
parity and charge superselection rules and additional symmetry
requirements \cite{EntanglementVsCorrelation2021,ding2022_faithful_entanglement,ding2023physical_entanglement}.
We will leverage these formulations to give explicit expressions
for the two-site symmetry-resolved REE, in the next Sections 
\ref{sec:sym-res-pure} and \ref{sec:sym-res-mixed}. Before proceeding,
we report some known inequalities between the REE, which for a 
two-site subsystem we will indicate as $E_{ij}$, the relative entropy
of correlations (mutual information, for two sites $I_{ij}$) and the
relative entropy of classical correlations (quantum discord, for two sites $Q_{ij}$). 
From the inclusion relationships between the 
correspondent sets, one has $I_{ij} \geq Q_{ij} \geq E_{ij}$.
One can also prove that $I_{ij} \geq Q_{ij} + C_{ij}$ where the 
\emph{classical correlation} $C_{ij}$ is defined as the QRE between
the closest classically correlated state and the closest uncorrelated
state, \ie the minimizers defining respectively $Q_{ij}$ and 
$I_{ij}$ 
\cite{Vedral2001,Discord&Classical_unified,Schilling_QuantumScience}.

\section{Symmetry-resolved entanglement}
\label{sec:sym-res}

In order to define our symmetry resolution of the two-site 
fermionic partial transpose and entanglement measures,
we first fix our representation of the two-site Fock space, 
density matrix and their symmetry properties.
For a single-band Hubbard model, the local Fock space has dimension 4
and is spanned by the holon $\ket{\bullet}$, spin-up $\ket{\uparrow}$, 
spin-down $\ket{\downarrow}$ and doublon $\ket{\uparrow\downarrow}$ states. 
Thus, for a system of two Hubbard sites the Fock space has dimension 
$4^2\equiv16$. Each of these 16 states can be labeled with an integer
$s=1,2,\dots,16$, such that the binary decomposition of the $(s-1)$ 
determines the occupations $n_{i\sigma}=0,1$ of the state with spin 
$\sigma=\uparrow,\downarrow$ at the site $i=1,2$. 
In our convention, the corresponding two-site Fock state reads
$\ket{s}=\ket{n_{1\uparrow},n_{2\uparrow},n_{1\downarrow},n_{2\downarrow}}$. 
The arbitrary choice of the order of the
occupation numbers is consistent with the convention in the \textsl{EDIpack}
library \cite{EDIpack2,EDIpack_code}, allowing for massively parallel 
diagonalization \cite{EDIpack,code_repo} in the numerical calculations we present 
in Sections \ref{sec:half_filled_MIT} and \ref{sec:doped_MIT}. More information
on our numerical implementation can be found in Appendix \ref{app:cdmft_ed}.

\begin{table}[ht]
\renewcommand{\arraystretch}{1.42}
    \centering
    \begin{minipage}{0.48\textwidth}
        \centering
        \begin{tabular}{c|l}
        $s$ &  $\ket{n_{1\uparrow},n_{2\uparrow},n_{1\downarrow},n_{2\downarrow}} \mapsto \ket{i} \otimes \ket{j}$ \\ \hline
        1   &  $  | 0,0,0,0\rangle\longmapsto     | \bullet \rangle\otimes | \bullet \rangle$\\
        2   &  $  | 1,0,0,0 \rangle\longmapsto    |\! \uparrow \rangle\otimes | \bullet \rangle$\\
        3   &  $  | 0,1,0,0 \rangle\longmapsto    | \bullet \rangle\otimes |\! \uparrow \rangle$\\
        4   &  $  | 1,1,0,0 \rangle\longmapsto    |\! \uparrow \rangle\otimes |\! \uparrow \rangle$\\
        5   &  $  | 0,0,1,0 \rangle\longmapsto    |\! \downarrow \rangle\otimes | \bullet \rangle$\\
        6   &  $  | 1,0,1,0 \rangle\longmapsto    |\! \uparrow  \downarrow \rangle\otimes | \bullet \rangle$\\
        7   &  $  | 0,1,1,0 \rangle\longmapsto    |\! \downarrow \rangle\otimes |\! \uparrow \rangle \times(-1)$\\
        8   &  $  | 1,1,1,0 \rangle\longmapsto    |\! \uparrow  \downarrow \rangle\otimes |\! \uparrow \rangle \times(-1)$
        \end{tabular}
    \end{minipage}
    \hfill
    \begin{minipage}{0.48\textwidth}
        \centering
        \begin{tabular}{c|l}
        $s$ &  $\ket{n_{1\uparrow},n_{2\uparrow},n_{1\downarrow},n_{2\downarrow}} \mapsto \ket{i} \otimes \ket{j}$ \\ \hline
        9   &  $  | 0,0,0,1 \rangle\longmapsto    | \bullet \rangle\otimes |\! \downarrow \rangle$\\
        10  &  $  | 1,0,0,1 \rangle\longmapsto    |\! \uparrow \rangle\otimes |\!  \downarrow \rangle$\\
        11  &  $  | 0,1,0,1 \rangle\longmapsto    | \bullet \rangle\otimes |\! \uparrow  \downarrow \rangle$\\
        12  &  $  | 1,1,0,1 \rangle\longmapsto    |\! \uparrow \rangle\otimes |\! \uparrow  \downarrow \rangle$\\
        13  &  $  | 0,0,1,1 \rangle\longmapsto    |\! \downarrow \rangle\otimes |\! \downarrow \rangle$\\
        14  &  $  | 1,0,1,1 \rangle\longmapsto    |\! \uparrow  \downarrow \rangle\otimes |\!  \downarrow \rangle$\\
        15  &  $  | 0,1,1,1 \rangle\longmapsto    |\! \downarrow \rangle\otimes |\! \uparrow  \downarrow \rangle \times(-1)$\\
        16  &  $  | 1,1,1,1 \rangle\longmapsto    |\! \uparrow  \downarrow \rangle\otimes |\! \uparrow  \downarrow \rangle \times(-1)$
        \end{tabular}
    \end{minipage}
    \caption{Fock states for a two-site fermionic system. As highlighted by the
    $\ket{i} \otimes \ket{j}$ notation, each of these basis states is separable in the
    two-site bipartition, and so are their classical mixtures. Hence, in this basis, a
    two-site density matrix $\rho_{ij}$ can be entangled only if off-diagonal elements 
    are present.}
    \label{tab:fock_basis}
\end{table}

Given the definition of two-site separable states (Eq.\,\ref{eq:sep}), it is
useful to reorder the occupation numbers in the form 
$\ket{n_{1\uparrow},n_{1,\downarrow}} \otimes \ket{n_{2\uparrow},n_{2,\downarrow}} \equiv \ket{i}\otimes\ket{j}$, 
so as to highlight that in this basis a diagonal two-site density matrix is 
separable in a classical mixture of products of single-site states. 
Accounting for fermionic anticommutation rules, the mapping is given as in 
Table\,\ref{tab:fock_basis}.

\smallskip

We highlight that in general the two considered sites can be either
isolated or embedded in a larger system. In the latter case, the 
$\ket{s}$ basis state only describes their contribution to the 
``whole'' basis state: $\ket{S}=\ket{s}\otimes\ket{s^\text{env.}}$, 
where $s^\text{env.}$ is the integer corresponding to the bit
decomposition of the degrees of freedom in the environment.
The two-site density matrix is then formally defined by a
partial trace over the environment degrees of freedom, as
\begin{equation}
(\rho_{ij})_{ss'}=
\sum_{s^\text{env.}}
\left(\bra{s}\otimes\bra{s^\text{env.}}\right) 
\rho 
\left(\ket{s'}\otimes\ket{s^\text{env.}}\right),
\qquad s,s'=1,\ldots,16,
\label{eq:2rdm}
\end{equation}
where $\rho$ describes the, possibly mixed, whole system.
The sum in Eq.\,\ref{eq:2rdm} contains $4^{|s^\text{env.}|}$ terms, 
where $|s^\text{env.}|$ indicates the total number of sites in the 
environment. Thus, the number of elements to sum is exponentially 
large in the environment dimension, requiring the use of fast
algorithms for implementing a direct numerical trace
(see, for instance, \cite{EDIpack2,EDIpack_code}). 

Remarkably, the Hubbard Hamiltonian Eq.~\ref{eq:hubbard} admits a 
symmetry group $G=U_\uparrow(1)\otimes U_\downarrow(1)$, which 
corresponds to the separate conservation of the total number of 
spin-up $N_\uparrow$ and spin-down $N_\downarrow$ electrons, \ie 
$[H,N_{\sigma}]=0$. As these conserved quantities are \emph{additive}
over subsystems, it can be proven that any reduced density matrix 
inherits an analogous conservation law for the corresponding reduced
quantum numbers $(n_\uparrow,n_\downarrow)$ \cite{bellomia_local_classical}.
For the two-site subsystem one has $n_\uparrow=n_{i\uparrow}+n_{j\uparrow}$, 
$n_\downarrow=n_{i\downarrow}+n_{j\downarrow}$ and $n_\sigma=0,1,2$. 
The associated block-diagonal decomposition of the two-site reduced 
density matrix is summarized in Table~\ref{tab:symmetry_sectors}.
Namely, the matrix elements in $\rho_{ij}$ that connect Fock states 
belonging to different symmetry sectors identically vanish. 
This symmetry-enforced simplification of $\rho_{ij}$ allows an easier
evaluation of the nonzero matrix elements, circumventing the need for
an explicit representation of the environment degrees of freedom
\cite{Roosz_2RDMfrom2GF,Bippus_Two-Site_Negativity_Hubbard}.
Furthermore, the symmetries of $\rho_{ij}$ and its partial transpose 
play a key role in classifying physically distinct origins of the 
two-site entanglement, as we will detail in the rest of the paper,
for both pure and mixed two-site states.  

\begin{table}[ht]
\centering
\renewcommand{\arraystretch}{1.2}
\begin{tabular}{c|c|c}
$(n_\uparrow,n_\downarrow)$ & Basis states ($s$) & Dimension \\ \hline
$(0,0)$ & (1) & 1 \\
$(1,0)$ & (2,3) & 2 \\
$(2,0)$ & (4) & 1 \\
$(0,1)$ & (5,9) & 2 \\
$(1,1)$ & (6,7,10,11) & 4 \\
$(2,1)$ & (8,12) & 2 \\
$(0,2)$ & (13) & 1 \\
$(1,2)$ & (14,15) & 2 \\
$(2,2)$ & (16) & 1
\end{tabular}
\caption{Decomposition of the two-site Fock space into symmetry sectors
labeled by the reduced conserved quantum numbers $(n_\uparrow,n_\downarrow)$.
The integer labels (s) refer to the basis states listed in Table~\ref{tab:fock_basis}.
The dimensions in the last column sum to 16.}
\label{tab:symmetry_sectors}
\end{table}

\subsection{Pure two-site states}
\label{sec:sym-res-pure}

\begin{figure}
    \centering
    \includegraphics[width=.7\linewidth]{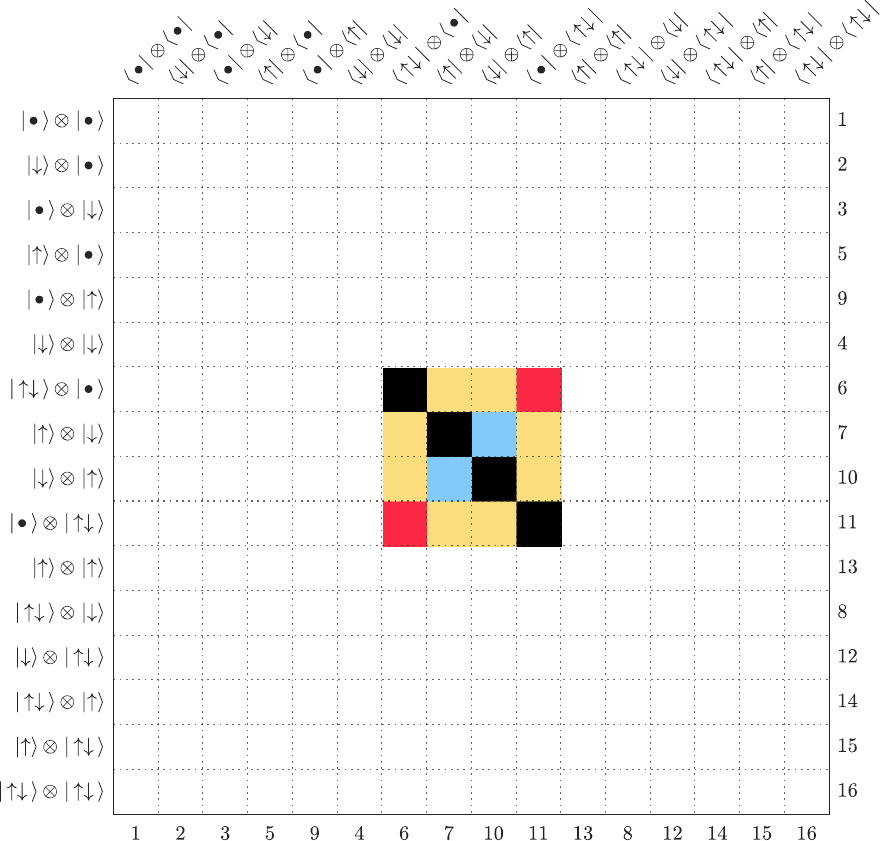}\\[2ex]
    \includegraphics[width=.7\linewidth]{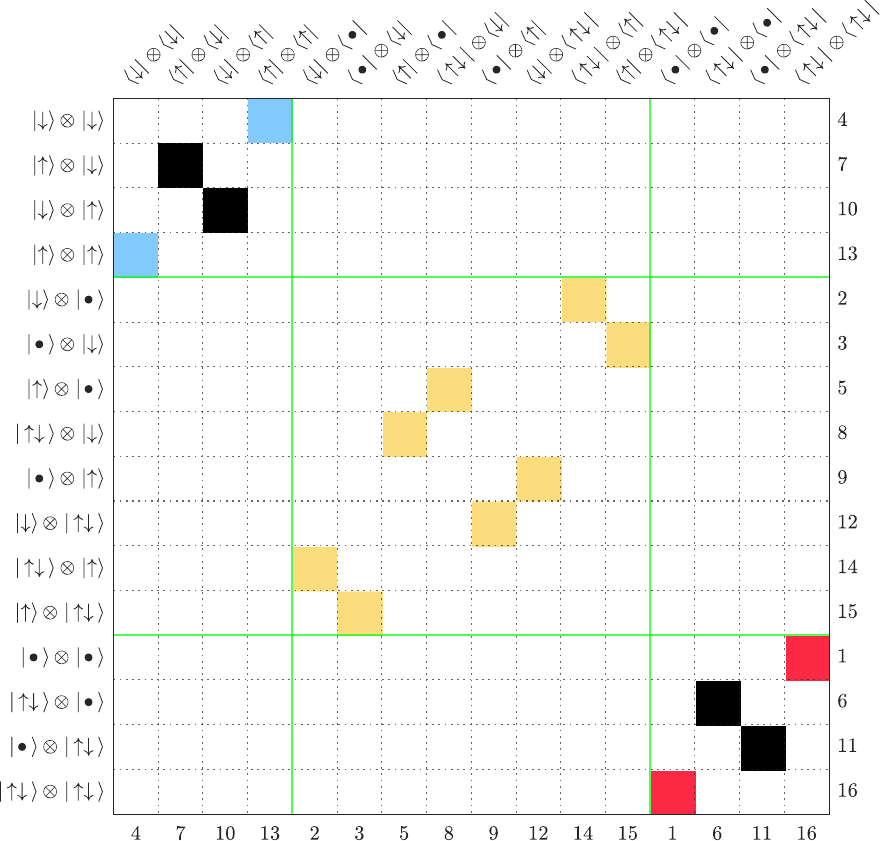}\\[1ex]
    \caption{Symmetry decomposition of the two-site 
    density matrix $\rho_{ij}$ (top panel) and its fermionic partial transpose
    $\rho^\mathrm{F}_{i(j)^\intercal}$ (bottom panel), for a pure Hubbard dimer,
    restricted to the half-filled sector. 
    The two-site entanglement $E_{ij}$ is introduced by
    the off-diagonal terms, which can be grouped into
    three symmetry classes. Hopping amplitudes (yellow),
    antiferromagnetic amplitudes (cyan) and holon-doublon amplitudes (magenta).
    White boxes represent zero elements.
    The Fock basis states are reordered to highlight the respective symmetry blocks
    in $\rho_{ij}$ and $\rho^\mathrm{F}_{i(j)^\intercal}$.}
    \label{fig:dimer_rdm_and_ppt}
\end{figure}

In Fig.\,\ref{fig:dimer_rdm_and_ppt} we schematically represent a pure 
two-site density matrix in the half-filled sector (top panel) and its 
fermionic partial transpose (bottom panel). 
White boxes represent matrix elements that vanish due to the conservation
of $n_\uparrow=1$ and $n_\downarrow=1$ on the dimer. Black boxes represent 
nonzero diagonal entries, each corresponding to a finite probability of 
populating a given Fock state.
Finally, colored boxes correspond to off-diagonal quantum amplitudes, 
which introduce entanglement between the two sites. The different
colors label the three different symmetry sectors (\ie diagonal blocks) in 
$\rho^\mathrm{F}_{i(j)^\intercal}$: cyan for amplitudes associated with 
no charge jump (a global spin-flip), yellow for amplitudes associated with
a 1-charge jump and magenta for amplitudes associated with a 2-charge jump 
(from two empty sites to two doubly-occupied sites, and vice versa). 
Please note that due to the ``space-time swap'' properties of the partial 
transpose \cite{PPT_as_space-time_swap}, in the original density matrix
$\rho_{ij}$ these amplitudes correspond respectively to antiferromagnetic
fluctuations (``$\color{xkcdCornflower}\uparrow\downarrow$''), 
hopping processes (``$\color{xkcdGold}t$'') and 
nonlocal holon-doublon (``{\color{xkcdStrawberry}hd}'') fluctuations.

\smallskip

Before proceeding to construct symmetry-resolved measures of the two-site entanglement,
based on this block-diagonal structure of the fermionic partial transpose, we highlight 
the connection between these three entanglement sectors and the celebrated 
\emph{superselection rules} for parity (P-SSR) and charge (N-SSR)
\cite{Wick1952,Wick1970,Bartlett2003,Cirac2004a,Cirac2004b,Dowling2006,SSR_lecturenotes}.
These constraints on the allowed quantum amplitudes in fermionic (sub)systems play a 
fundamental role in constructing a well-defined quantum information theory
of indistinguishable fermions
\cite{Banhuls2007,Friis2013,Friis2016,Benatti2020,Szalay2021}
and have recently been applied with great success in quantum chemistry
\cite{DissotiationParadoxSchilling,EntanglementVsCorrelation2021,Schilling_QuantumScience,ding2022_faithful_entanglement,ding2023physical_entanglement,Faraday_Natural_Orbitals} 
and lattice models for strongly-correlated materials
\cite{MonkmanSirker2020,Galler_SSR,DelMaestro_SSR_spinless,DelMaestro_SSR_attractive,Bellomia2023,BellomiaPhD,Canio_orbital_concurrence,thermal_death_ssr,SSR_PG_ColdAtoms}.

The P-SSR expresses a fundamental constraint of fermionic quantum mechanics, 
imposing that any operation in a given system or subsystem must not 
entail quantum superpositions of states with different parity of the 
fermion number, as these states cannot be prepared or manipulated 
\cite{Wick1952}. Consequently, in a bipartite two-site setup, neither 
Alice nor Bob can access single-site states that violate the P-SSR, 
for any meaningful quantum information processing task, even when these
superpositions are encoded in the two-site density matrix $\rho_{ij}$.
Failing to enforce the P-SSR restriction notoriously leads to mathematical
inconsistencies \cite{Friis2013,Benatti2020,Szalay2021},
violations of relativistic causality (\ie the no-signaling theorem) 
\cite{Friis2016}, and substantial brittleness of the predicted entanglement
against time evolution in typical quantum protocols \cite{Fate_of_Entanglement}.
Hopping amplitudes connect states with different local parity and therefore 
violate the P-SSR when a two-site bipartition of $\rho_{ij}$ is considered. 
Hence, the hopping sector in the fermionic partial transpose, while 
contributing a portion of the full fermionic entanglement, corresponds
to quantum correlations that are not \emph{fundamentally accessible} in 
any quantum information processing task. For this reason, the hopping
entanglement does not constitute a genuine quantum resource.

In addition to the P-SSR, for systems that conserve the global 
fermion number, as the Hubbard model considered in this work, one 
can further consider the more restrictive N-SSR \cite{Wick1970}.
As for the fermion parity, one can prove that the global charge 
conservation is inherited as a local charge conservation in the
single-site reduced density matrix \cite{bellomia_local_classical},
further restricting the allowed local operations on the two subsystems.
Hence, in the absence of any charge-non-conserving perturbation, 
the \emph{operationally accessible} portion of entanglement
in two-site Hubbard (sub)systems excludes also the holon-doublon sector
\cite{ding2023physical_entanglement,Bellomia2023,BellomiaPhD}.

All in all, we have a clear-cut correspondence between symmetry-resolved,
full and superselected two-site entanglement, as
\begin{equation}
\begin{cases}
E^\text{N-SSR}_{ij}\!\!\!\!\!\!&\equiv\, E^{\uparrow\downarrow}_{ij} \\[1ex]
E^\text{P-SSR}_{ij}\!\!\!\!\!\!&\equiv\, E^{\uparrow\downarrow}_{ij} 
                    \,\overset{\star}{+}\,E^\mathrm{hd}_{ij} \\[1ex]
E^\text{(full)}_{ij}\!\!\!\!\!\!&\equiv\, E^{\uparrow\downarrow}_{ij} 
                    \,\overset{\star}{+}\,E^\mathrm{hd}_{ij}
                    \,\overset{\star}{+}\,E^{t}_{ij}
\end{cases} \label{eq:ssr_connection}
\end{equation}
where the star above the ``+'' indicates that the specific composition rule 
across sectors will depend on the considered entanglement measure. As we 
will see in the following, the logarithmic negativity is composed through
exponentiation inside the logarithm, while the REE is truly additive over
the symmetry sectors, due to the properties of the quantum relative entropy.

\smallskip 

Having clarified the general relationship between the symmetry resolution
of the fermionic partial transpose and charge/parity superselection rules,
we introduce the three symmetry-resolved negativities by applying the PPT
criterion on each block.
As the entanglement negativity (Eqs.\,\ref{eq:negativity_svd} and 
\ref{eq:fermionic_negativity}) is constructed by summing over singular values,
and the SVD of a block-diagonal matrix is the union of the singular values
for each block, we can meaningfully define a negativity for each symmetry
block. The full negativity is then given as 
\begin{equation}
    \mathcal{N}_{ij}^\mathrm{F} = \log\left[
    \exp(\mathcal{N}_{ij}^{\uparrow\downarrow}) +
    \exp(\mathcal{N}_{ij}^{\mathrm{F}t}) +
    \exp(\mathcal{N}_{ij}^{\mathrm{hd}}) - 2 \right].
\end{equation}
A careful inspection of Eq.\,\ref{eq:fermionic_ppt_phase} reveals how the
special fermionic phases $\phi^{\lambda\nu}_{ij}$ 
vanish for both the antiferromagnetic and the holon-doublon sectors. Hence,
$\mathcal{N}_{ij}^{\uparrow\downarrow}$ and $\mathcal{N}_{ij}^{\mathrm{hd}}$
can be computed with the standard ``bosonic'' transpose, \ie by summing over the
negative eigenvalues of $\rho_{i(j)^\intercal}$, in the respective sectors
(see Eq.\,\ref{eq:negativity_eig}).
Given the particularly sparse nature of the considered pure state, it follows
that antiferromagnetic entanglement $E^{\uparrow\downarrow}_{ij}$ is always 
present for any finite 
$\ketbra{\uparrow}{\downarrow}\otimes\ketbra{\downarrow}{\uparrow}$ 
amplitude, as there are no corresponding diagonal elements in 
$\rho_{i(j)^\intercal}^\mathrm{F}$ (see bottom panel in 
Fig.\,\ref{fig:dimer_rdm_and_ppt}). 
Similarly, the holon-doublon entanglement $E^\mathrm{hd}_{ij}$ is
guaranteed by any finite
$\ketbra{\uparrow\downarrow}{\bullet}\otimes\ketbra{\bullet}{\uparrow\downarrow}$
amplitude.
In particular one has
\begin{align}
    \mathcal{N}_{ij}^{\uparrow\downarrow} \Big|_\text{pure} 
    = \log(1 + 2\left|\rho_{ij}(7,10)\right|)\,, \qquad
    \mathcal{N}_{ij}^\mathrm{hd} \Big|_\text{pure} 
    = \log(1 + 2\left|\rho_{ij}(6,11)\right|)\,.
\end{align}
In the 1-charge jump sector of the fermionic partial transpose, the 
$\phi^{\lambda\nu}_{ij}$ factors are instead nontrivial, so that the 
hopping negativity $\mathcal{N}_{ij}^{\mathrm{F}t}$ must be computed 
from the corresponding singular values. 
For this reason we keep the ``F'' superscript, alongside the hopping
symbol (``$t$'').

\smallskip

Before turning to the quantum relative entropy approach to quantify
entanglement as a distance from the separable set, we briefly 
comment on the sizes of the three symmetry sectors in the partial
transpose. The two superselected sectors
($\uparrow\downarrow$ and hd) are small enough for the
reverse PPT criterion to hold \cite{Horodecki97}, so that a vanishing 
negativity in the antiferromagnetic sector faithfully detects zero 
N-SSR entanglement and a further vanishing holon-doublon negativity 
ensures that the state is separable under P-SSR.
Hence, the presence of accessible two-site entanglement is exactly
captured by the two SSR-compliant symmetry-resolved components.
On the other hand, the hopping sector has dimension 
$8>2\times3$, therefore we cannot exclude the existence of bound
entanglement that cannot be detected by the PPT criterion. 
Remarkably, as we will see in detail, in this work the hopping
negativity never vanishes, for any model parameter in both the
analytical results for the isolated Hubbard dimer and the numerical
simulations of the two-dimensional Hubbard model.
It follows that, at most, we might be underestimating the hopping
contribution to the two-site entanglement, but never fully neglecting
its existence.

\smallskip

Finally, we discuss  how to quantify entanglement through the QRE approach.
As discussed in Section \ref{sec:entanglement}, the geometrical optimization problem 
that underlies the definition of the relative entropy of entanglement is trivially 
solved for pure bipartite systems. Hence, for a pure dimer we can quantify $E_{ij}$
exactly, by computing the single-site von Neumann entropy $S_i$ 
\cite{Franca_LDAentropy,XiDai2013,Cocchi_ColdAtoms,WalshLetter2019,WalshPRXQuantum2020,Franca_local_entropy_superc}.
On the other hand, enforcing superselection rules inherently maps quantum superpositions 
of states with different charge/parity into their classical mixture counterparts,
as the off-diagonal amplitudes connecting the different charge/parity sectors are removed. 
Because of this, minimizing the QRE under N-SSR or P-SSR on a pure
state is in general as hard as for mixed states \cite{REE_NPcomplete,Discord_NPcomplete}.
Nonetheless, for the special case of measuring the
entanglement between a single fermionic orbital and
the rest of the whole, closed (pure) system, closed
formulas have been derived 
\cite{EntanglementVsCorrelation2021}.
Most remarkably, these expressions for the REE under
P-SSR and N-SSR depend only on the single-site reduced
density matrix, as for the von Neumann entropy $S_i$.
Given that here our system is an isolated pure dimer,
these expressions can be applied to evaluate the two-site
REE in the superselected sectors.
Setting $p_\bullet = \expval{\ketbra{\bullet}{\bullet}} = 
\expval{1 - n_i + n_{i\uparrow}n_{i\downarrow}}$,
$p_\uparrow = \expval{\ketbra{\uparrow}{\uparrow}} = 
\expval{n_{i\uparrow} - n_{i\uparrow}n_{i\downarrow}}$,
$p_\downarrow = \expval{\ketbra{\downarrow}{\downarrow}} = 
\expval{n_{i\downarrow} - n_{i\uparrow}n_{i\downarrow}}$,
$p_{\uparrow\downarrow} = 
\expval{\ketbra{\uparrow\downarrow}{\uparrow\downarrow}} = 
\expval{n_{i\uparrow}n_{i\downarrow}}$, one has
(see \cite{EntanglementVsCorrelation2021} for a derivation)
\begin{align}
    E_{ij}^\text{N-SSR}\Big|_\text{pure}  &=\,\,
    (p_\uparrow + p_\downarrow) 
    \log(p_\uparrow + p_\downarrow)
    - p_\uparrow \log(p_\uparrow)
    - p_\downarrow \log(p_\downarrow), \label{eq:nssr_ree_pure}\\[1ex]
    E_{ij}^\text{P-SSR}\Big|_\text{pure}  &=\,\, E_{ij}^\text{N-SSR}\Big|_\text{pure}  +\,\,
    (p_\bullet + p_{\uparrow\downarrow}) 
    \log(p_\bullet + p_{\uparrow\downarrow})
    - p_\bullet \log(p_\bullet)
    - p_{\uparrow\downarrow}\log(p_{\uparrow\downarrow}). \label{eq:pssr_ree_pure}
\end{align}

We readily see how the antiferromagnetic and the holon-doublon sectors are composed additively to obtain the REE
under P-SSR. Given that
\begin{equation}
    E_{ij}^\text{(full)}\Big|_\text{pure} =\,\, S_i =
    - p_\bullet \log(p_\bullet)
    - p_\uparrow \log(p_\uparrow)
    - p_\downarrow \log(p_\downarrow)
    - p_{\uparrow\downarrow}\log(p_{\uparrow\downarrow}),
\end{equation}
the relationship between symmetry resolution and 
superselection rules (Eq.\,\ref{eq:ssr_connection}), 
in the framework of quantum relative entropies, 
is realized as
\begin{equation}
\begin{aligned}
    E_{ij}^{\uparrow\downarrow}\Big|_\text{pure} &=\,\,
    E_{ij}^\text{N-SSR}\Big|_\text{pure}\,\,,  \qquad
    E_{ij}^\mathrm{hd}\Big|_\text{pure} =\,\,
    E_{ij}^\text{P-SSR}\Big|_\text{pure} -\,\, 
    E_{ij}^\text{N-SSR}\Big|_\text{pure}\,\,, \\[1ex]
    E_{ij}^{t}\Big|_\text{pure} &=\,\,
    -(p_\uparrow + p_\downarrow) 
    \log(p_\uparrow + p_\downarrow)
    -(p_\bullet + p_{\uparrow\downarrow}) 
    \log(p_\bullet + p_{\uparrow\downarrow}). \label{eq:sym_res_ree_pure}
\end{aligned}
\end{equation}
This reproduces a known splitting of the entanglement entropy of pure states
into \emph{configurational} and \emph{number entropy} parts \cite{Lukin2019_MBL,MonkmanSirker2020,MonkmanSirker2023}.
In this construction $E_{ij}^{\uparrow\downarrow}$ is the configurational part, while $E_{ij}^\mathrm{hd}+E_{ij}^{t}$ is the
Shannon entropy of the local charge distribution (\ie the number entropy).
The $E_{ij}^{t}$ term similarly corresponds to the Shannon entropy of
the local charge parity.

\subsection{Mixed two-site states}
\label{sec:sym-res-mixed} 

\begin{figure}
    \centering
    \includegraphics[width=.7\linewidth]{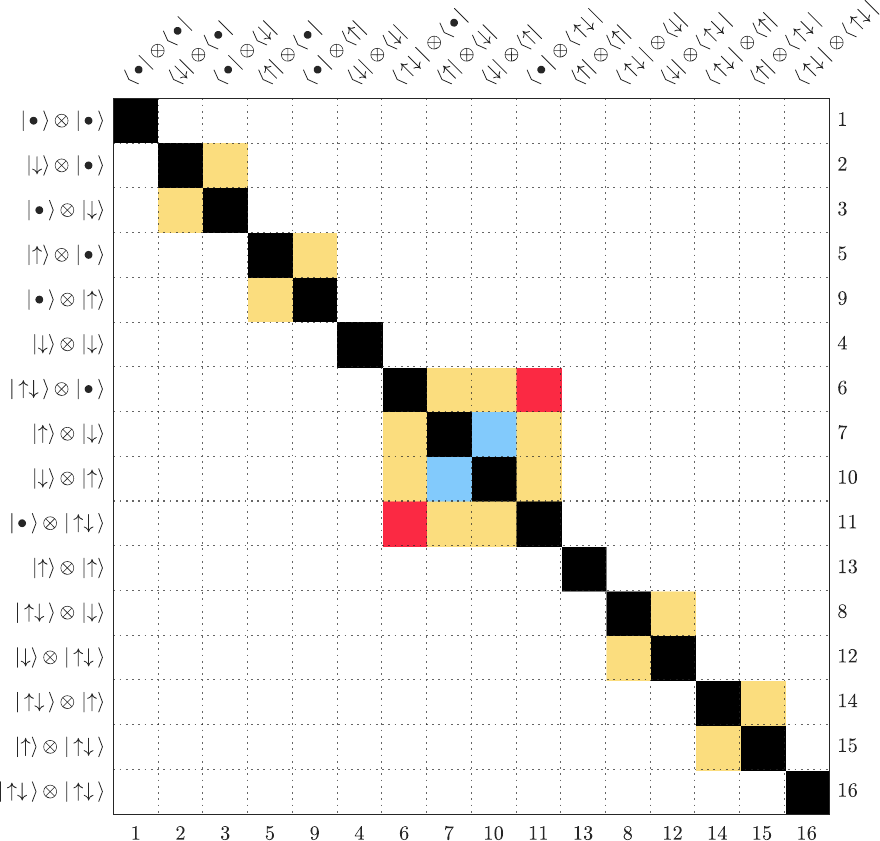}\\[2ex]
    \includegraphics[width=.7\linewidth]{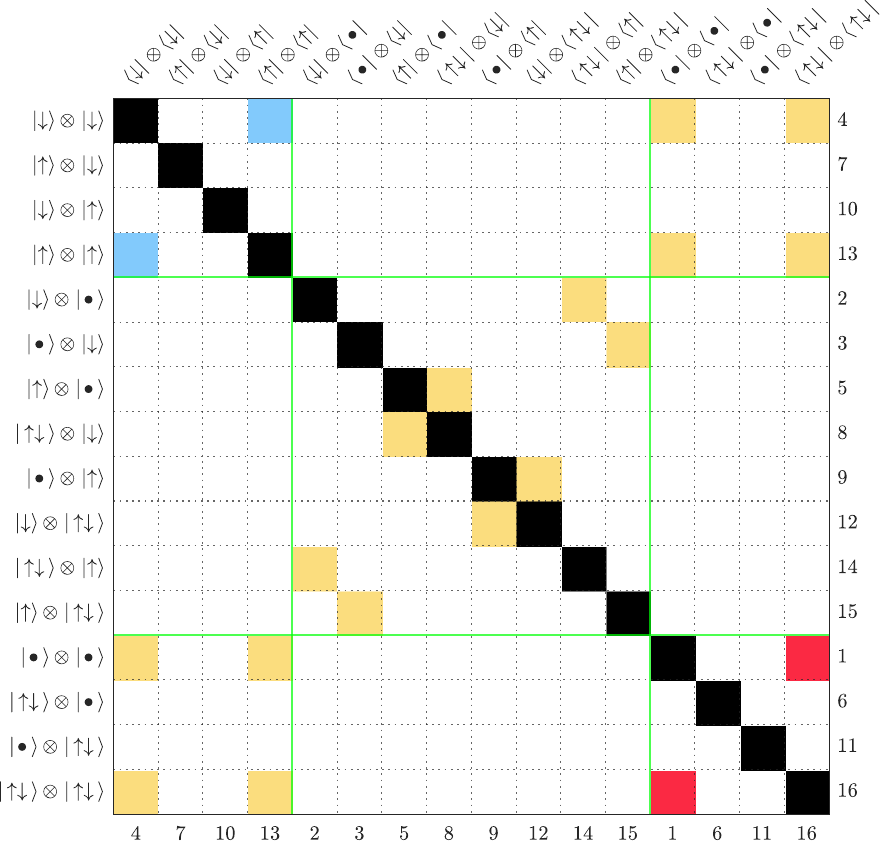}\\[1ex]
    \caption{Symmetry decomposition of a generic, mixed two-site 
    density matrix $\rho_{ij}$ (top panel) and its fermionic partial transpose
    $\rho^\mathrm{F}_{i(j)^\intercal}$ (bottom panel), for a Hubbard system. 
    The mixed character of the state populates all the charge symmetry sectors in
    $\rho_{ij}$. The resulting additional hopping amplitudes break the symmetry 
    decomposition of $\rho^\mathrm{F}_{i(j)^\intercal}$, unless they are removed 
    by the parity superselection rule. White boxes represent zero elements.
    The Fock basis states are reordered as in Fig.\,\ref{fig:dimer_rdm_and_ppt}, to highlight the 
    symmetry blocks in $\rho_{ij}$ and $\rho^\mathrm{F}_{i(j)^\intercal}$.}
    \label{fig:mixed_rdm_and_ppt}
\end{figure}

In the previous section,
we have established a symmetry resolution of 
entanglement in pure two-site states, in terms
of a block structure in the corresponding fermionic
partial transpose.
We have also provided a direct
connection between two of the three entanglement 
terms to charge and parity superselection rules,
as well as provided a clear physical origin to each
contribution: spin-singlet entanglement, holon-doublon
entanglement and hopping entanglement.
In this section we move forward to consider 
\emph{mixed} two-site states, how the block 
structure of the corresponding fermionic partial 
transpose changes and how a symmetry resolution of
entanglement can be built from it.

In the top panel of Fig.\,\ref{fig:mixed_rdm_and_ppt}
we represent the most general mixed two-site state,
consistent with the symmetries of a Hubbard Hamiltonian:
as all charge sectors are populated, we get a full diagonal
and four additional amplitudes, other than the elements
already discussed for the half-filled sector, in Section
\ref{sec:sym-res-pure}. These additional off-diagonal terms
are associated to single-electron hopping processes, hence
forbidden by the parity superselection rule. 

In the fermionic partial transpose (bottom panel in 
Fig.\,\ref{fig:mixed_rdm_and_ppt}), the additional hopping
amplitudes end up connecting the antiferromagnetic and the
holon-doublon sectors, hindering a clean separation of the
three sources of entanglement, in both the negativity and 
the relative entropy of entanglement. 
Yet, under P-SSR, which we recall to be rooted in 
\emph{fundamental} constraints to the superpositions allowed
on each subsystem, all hopping (yellow) off-diagonal elements
are discarded \cite{ding2022_faithful_entanglement,ding2023physical_entanglement}.
Hence, even for a mixed two-site state, the parity superselected
entanglement can be resolved into a spin-singlet (0-charge jump,
or charge superselected) and a holon-doublon (2-charge jump)
component. In contrast to pure two-site states, a true 1-charge
jump sector is not well defined, of course, but we can 
uniquely identify a hopping entanglement as the difference 
between the full two-site entanglement and the P-SSR portion of it.
Namely, we enforce the additive structure
(Eq.\,\ref{eq:ssr_connection}) that holds for pure states,
also to mixed states.

\smallskip

As far as the superselected negativities are concerned, we note 
that also in the mixed case the two off-diagonal elements are
devoid of any fermionic partial transpose phase 
(Eq.\,\ref{eq:fermionic_ppt_phase}) and the dimension of the
relevant blocks in $\rho_{i(j)^\intercal}$ is small enough to
allow the reverse PPT criterion to hold. Hence, once again, the
absence of any negative eigenvalue in each of these sectors strictly
proves the absence of spin-singlet and/or holon-doublon entanglement.
Unlike the pure case, we highlight that we have a full diagonal
in each sector, so the mere presence of a finite off-diagonal
term does not prove the presence of superselected entanglement.
The PPT condition must be explicitly assessed by comparing
the magnitude of diagonal and off-diagonal terms, to check
\gb{for a possible change of sign in the smaller eigenvalue.
Each superselected block of $\rho^\mathrm{F}_{i(j)^\intercal}$ is
a $2\times2$ Hermitian matrix, whose trace is always nonnegative,
so that at most one of its two eigenvalues can be negative
and this happens if and only if its determinant is negative.
Hence, superselected entanglement is present in each sector if and only if
\begin{align}
    \bigl|\rho_{ij}(7,10)\bigr|^2 &> \rho_{ij}(4,4)\,\rho_{ij}(13,13)\,,
    \enspace\text{for 0-charge jump}, \label{eq:nssr_condition}\\
    \bigl|\rho_{ij}(6,11)\bigr|^2 &> \rho_{ij}(1,1)\,\rho_{ij}(16,16)\,,
    \enspace\text{for 2-charge jump}, \label{eq:pssr_condition}
\end{align}
namely each superselected off-diagonal amplitude must exceed 
the geometric mean of the two diagonal populations it competes
with.}\\
\gb{Furthermore, the nonnegativity of $\rho_{ij}$ bounds
each off-diagonal amplitude by the two populations it connects,
\ie
$|\rho_{ij}(6,11)|^2 \leq \rho_{ij}(6,6)\,\rho_{ij}(11,11)$ and
$|\rho_{ij}(7,10)|^2 \leq \rho_{ij}(7,7)\,\rho_{ij}(10,10)$.
Holon-doublon entanglement thus requires
$\rho_{ij}(6,6)\,\rho_{ij}(11,11) > \rho_{ij}(1,1)\,\rho_{ij}(16,16)$.
In other words, neighboring sites must host a coherent 
holon-doublon pair more often than two holons or two doublons,
\ie beyond what independent occupations would allow.
Analogously, spin-singlet entanglement requires coherent 
antiparallel spin configurations to be favored over parallel ones.}
Leveraging the $\mathrm{SU}(2)$ symmetry that holds for the Hubbard
systems we are interested in gives an even clearer physical meaning to the antiferromagnetic entanglement condition. The 0-charge jump amplitude can be 
expressed in the diagonal basis as 
$\rho_{ij}(7,10) = \rho_{ij}(7,7)-\rho_{ij}(4,4)$, and one further
as $\rho_{ij}(4,4) = \rho_{ij}(13,13)$ and 
$\rho_{ij}(7,7) = \rho_{ij}(10,10)$. So, Eq.\,\ref{eq:nssr_condition}
simplifies to $\rho_{ij}(7,7) > 2\rho_{ij}(4,4)$. 
Namely, one must have a mixing weight for the spin-singlet state 
$\left(\ket{\uparrow}\otimes\ket{\downarrow} - 
\ket{\downarrow}\otimes\ket{\uparrow}\right)/\sqrt{2}$,
larger than the sum of the weights
of the three spin-triplet states $\ket{\uparrow}\otimes\ket{\uparrow}$,
$\ket{\downarrow}\otimes\ket{\downarrow}$ and 
$\left(\ket{\uparrow}\otimes\ket{\downarrow} + 
\ket{\downarrow}\otimes\ket{\uparrow}\right)/\sqrt{2}$
\cite{ding2022_faithful_entanglement,SSR_PG_ColdAtoms}.
It follows that a classical mixing of spin-singlet and spin-triplet 
states can destroy the antiferromagnetic entanglement, if this
threshold is exceeded. We will show in Sections \ref{sec:dimer} and 
\ref{sec:cdmft_results} how the embedding of a Hubbard dimer in
a lattice does indeed destroy spin-singlet entanglement, as long as 
the system is in a weakly-interacting or overdoped regime.

\smallskip

Moving to the REE framework, we can quantify the entanglement of the
spin-singlet and holon-doublon sectors leveraging once again some closed
formulas proposed by Ding et al. 
\cite{ding2022_faithful_entanglement,ding2023physical_entanglement}, 
under P-SSR and N-SSR plus some additional symmetry constraints. 
Their validity is tied to the conservation of charge $N$, the total
spin $\vec{S}^2$ and its $S_z$ component in the given two-site density
matrix. Furthermore, the two given sites $i$ and $j$ must be 
interchangeable. Unfortunately, these constraints exclude many 
symmetry-broken phases.
Yet, for various 
paramagnetic states of the Hubbard model these formulas hold
and have been employed to study superselected entanglement across the
Mott transition in plaquette CDMFT \cite{Bellomia2023}, as well as for 
the analytical investigation of thermal entanglement and quench dynamics
in isolated Hubbard dimers \cite{Zavatti_Magic_Hubbard_dimer}.

Referring to \cite{ding2022_faithful_entanglement} for a full derivation,
we here report the closed formulas for the entanglement in the 
antiferromagnetic and the holon-doublon sectors, respectively
$\mathbb{S}^{\uparrow\downarrow}$ = Span\{\,$\ket{\Psi_4}$, $\ket{\Psi_7}$, $\ket{\Psi_{10}}$, $\ket{\Psi_{13}}$\,\}
and $\mathbb{S}^\mathrm{hd}$ = Span\{\,$\ket{\Psi_1}$, $\ket{\Psi_6}$, $\ket{\Psi_{11}}$, $\ket{\Psi_{16}}$\,\}, as
\begin{equation}
    E_{ij}^{\uparrow\downarrow}
    =\sum_{n\in\mathbb{S}^{\uparrow\downarrow}} 
      \nu_n \log{\frac{\nu_n}{\mu_n}}, \qquad
    E_{ij}^\mathrm{hd}
    =\sum_{n\in\mathbb{S}^\mathrm{hd}} 
      \nu_n \log{\frac{\nu_n}{\mu_n}},
      \label{eq:sym_res_ree_mixed}
\end{equation}
where
\begin{align*}
    \mu_{7} = \frac{b_1 + c_1 + \sqrt{d_1}}{4(a_1-\nu_{10})},\,\,
    \mu_{10} = \frac{b_1 - c_1 - \sqrt{d_1}}{4(a_1-\nu_{7})},\,\,
    &\mu_{6}  = \frac{b_2 + c_2 + \sqrt{d_2}}{4(a_2-\nu_{11})},\,\,
    \mu_{11} = \frac{b_2 - c_2 - \sqrt{d_2}}{4(a_2-\nu_{6})},\\[1mm]
    \mu_{4} = \nu_{4} + \frac{\nu_{7}+\nu_{10}-\mu_{7}-\mu_{10}}{2},\quad
    &\mu_{13} = \nu_{13} + \frac{\nu_{7}+\nu_{10}-\mu_{7}-\mu_{10}}{2},\\[1mm]
    \mu_{1} = \nu_{1} + \frac{\nu_{6}+\nu_{11}-\mu_{6}-\mu_{11}}{2},\quad
    &\mu_{16} = \nu_{16} + \frac{\nu_{6}+\nu_{11}-\mu_{6}-\mu_{11}}{2} ,
\end{align*}
in terms of
\begin{align*}
    &\nu_1 = \rho_{ij}(1,1), \quad
    \nu_4 = \rho_{ij}(4,4), \quad
    \nu_{13} = \rho_{ij}(13,13), \quad
    \nu_{16} = \rho_{ij}(16,16), \\
    &\!\!\!\!\!\begin{pmatrix}
        \lambda_6 & 0 \\
        0 & \lambda_{11}
    \end{pmatrix} = 
    R^{\dagger}
    \begin{pmatrix}
        \rho_{ij}(6,6) & \rho_{ij}(6,11)\\
        \rho_{ij}(11,6) & \rho_{ij}(11,11)
    \end{pmatrix}
    R,\\[2mm]
    &\!\!\!\!\!\begin{pmatrix}
        \lambda_7 & 0\\
        0 & \lambda_{10}
    \end{pmatrix} = 
    R^{\dagger}
    \begin{pmatrix}
        \rho_{ij}(7,7) & \rho_{ij}(7,10) \\
        \rho_{ij}(10,7) & \rho_{ij}(10,10)
    \end{pmatrix}
    R, \quad \text{with}\enspace R =
    \begin{pmatrix}
        1 & -1 \\
        1 & +1
    \end{pmatrix} \times \frac{1}{\sqrt2}, \\
    &\nu_{7} = \max(\lambda_7,\lambda_{10}), \quad
    \nu_{10} = \min(\lambda_7,\lambda_{10}), \\
    &\nu_{6} = \max(\lambda_6,\lambda_{11}), \quad
    \nu_{11} = \min(\lambda_6,\lambda_{11}), \\
    &a_1 = \nu_{7} + \nu_{10} + \nu_{4} + \nu_{13}, \quad 
    b_1 = a_1^2 - (\nu_{4} - \nu_{13})^2, \quad
    c_1 = (\nu_{7} - \nu_{10})a_1, \\
    &a_2 =  \nu_{1} + \nu_{6} + \nu_{11} + \nu_{16}, \quad
    b_2 = a_2^2 - (\nu_{1} - \nu_{16})^2, \quad
    c_2 = (\nu_{6} - \nu_{11})a_2, \\
    &d_1 = (\nu_{4}+\nu_{13})^2(\nu_{7}-\nu_{10})^2
           +8\nu_{4}\nu_{13}(2\nu_{4}\nu_{13}
           +(\nu_{4}+\nu_{13})(\nu_{7}-\nu_{10})
           +2\nu_{7}\nu_{10}), \\
    &d_2 = (\nu_{1}+\nu_{16})^2(\nu_{6}-\nu_{11})^2
           +8\nu_{1}\nu_{16}(2\nu_{1}\nu_{16}
           +(\nu_{1}+\nu_{16})(\nu_{6}-\nu_{11})
           +2\nu_{6}\nu_{11}). \\
\end{align*}

Once again, the antiferromagnetic sector generates the N-SSR entanglement,
while the P-SSR entanglement is given by summing over both 
$\mathbb{S}^{\uparrow\downarrow}$ and $\mathbb{S}^\mathrm{hd}$
\cite{ding2022_faithful_entanglement}, 
given the additive properties of relative entropies.
We note that, contrary to the negativity case, there is no feasible way
to evaluate the REE beyond superselection rules, so we cannot access the
hopping entanglement $E_{ij}^t$, in the mixed case. 
In Sections \ref{sec:dimer} and \ref{sec:cdmft_results} we will then rely
only on $\mathcal{N}_{ij}^{\mathrm{F}t}$ for quantifying it.

\subsection{Analytical testbed: isolated and noninteracting systems}
\label{sec:dimer}

To illustrate the main features of the symmetry-resolved negativity and REE,
in this section we will derive exact analytical results for the ground state
of an isolated Hubbard dimer, at half-filling and varying interaction strength.
In the noninteracting limit, we will also discuss the effects of mixing when 
the dimer is embedded in an infinite square lattice. In this way, we showcase
the properties of symmetry-resolved negativity and REE for both pure and
mixed two-site states, in a simple exact testbed, before diving into numerical
results for the two-dimensional Hubbard model, obtained from cellular dynamical
mean-field theory.

\subsubsection{Ground state of an isolated Hubbard dimer}

We consider an isolated system formed by two Hubbard sites with arbitrary interaction strength $U$,
coupled by a hopping amplitude $t$. 
In terms of the $\ket{i}\otimes\ket{j}$ Fock basis introduced in Section~\ref{sec:sym-res}
the ground state at half-filling is analytically given as
\begin{equation}
    \mid\!\!\text{GS} \rangle = 
    N \left[\mid\uparrow\downarrow\rangle \otimes \mid\!\!\bullet\rangle+
    \frac{E_+}{2t} \Bigl(
    \mid\uparrow\rangle \otimes \mid\downarrow\rangle -
    \mid\downarrow\rangle \otimes \mid\uparrow\rangle \Bigr) +
    \mid\!\!\bullet\rangle \otimes \mid\uparrow\downarrow\rangle
    \right], \label{eq:dimer_gs}
\end{equation}
with 
$$
N = \left[2 + \frac{1}{2t^2} 
\left(\frac{U+\Delta}{2}\right)^2\right]^{-\frac{1}{2}},
$$
$$
E_\pm = \frac{U \pm \Delta}{2}, \qquad 
\Delta = \sqrt{16t^2+U^2}.
$$

We note that $E_-$ is the ground state energy and 
$\mathcal{D} = N^2$ is the double occupancy.
The corresponding ground state density matrix is given by the projector
$\rho =\, |\text{GS} \rangle \!\langle\text{GS}|$.
Being $\rho$ pure, the full entanglement between the two sites $i$ and $j$
is uniquely defined by the single-site reduced density matrix
$$
\rho_i = \mathrm{Tr}_j \rho = \rho_j = \mathrm{Tr}_i \rho
= N^2 \Bigl( 
      \mid\!\!\bullet\rangle\langle \bullet\!\!\mid +
      \mid\!\!{\uparrow\downarrow}\rangle\langle {\uparrow\downarrow}\!\!\mid
   \Bigr) +
   \left(\frac{1}{2}-N^2\right) \Bigl(
      \mid\uparrow\rangle\langle \uparrow\mid +
      \mid\downarrow\rangle\langle \downarrow\mid
   \Bigr) 
$$

In particular, the QRE distance from the set of two-site separable states,
\ie the REE, is simply given as the von Neumann entropy of the single-site RDM
\begin{equation}
    E_{ij} = S_i = S_j = -2N^2\log(N^2) - 2\left(\frac{1}{2}-N^2\right)\log(\frac{1}{2}-N^2).
\end{equation}

As for pure states all quantum and classical correlations are fully determined by the entanglement,
also the QRE distance from the set of two-site product states, \ie the mutual information,
is fully determined by the single-site von Neumann entropy. In particular, one can explicitly prove
that quantum and classical correlations are additive and equipartite in pure states
$I_{ij} = Q_{ij} + C_{ij} = 2E_{ij}$, as $Q_{ij} = C_{ij} = E_{ij}$ 
\cite{Vedral2001,Discord&Classical_unified,Schilling_QuantumScience}.
This is a special case of the general inequality $I_{ij} \geq E_{ij}$, relating the mutual 
information $I_{ij}$ -- which we recall to give an upper bound to all correlation functions -- and 
the entanglement $E_{ij}$. 

\begin{figure}
    \centering
    \includegraphics[width=0.95\linewidth]{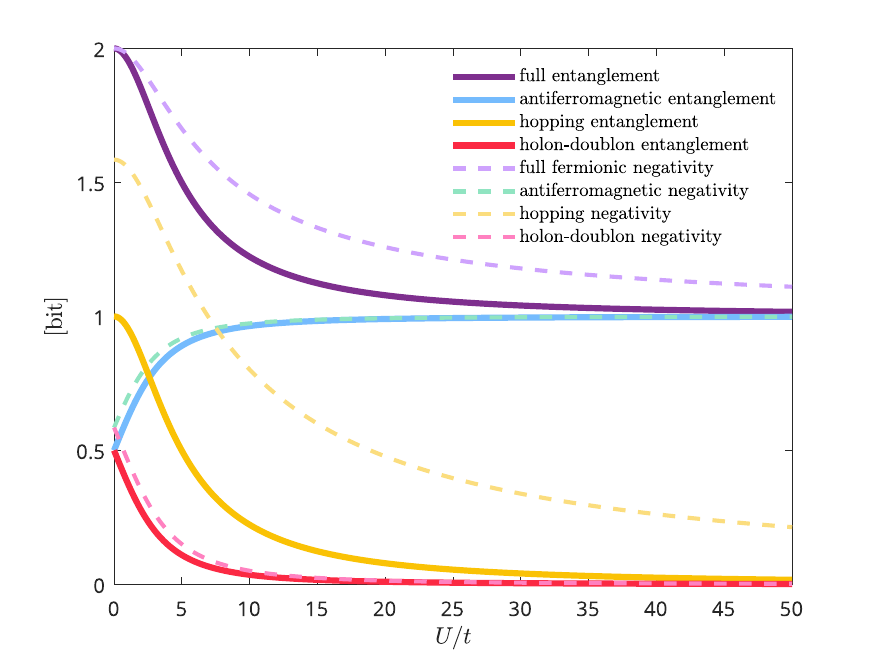}
    \caption{Different measures of full and symmetry-resolved entanglement between the two sites
    constituting an isolated Hubbard dimer, in its half-filled ground state.}
    \label{fig:dimer}
\end{figure}

\smallskip

Following the construction given in Section~\ref{sec:negativity}, 
we can evaluate the two-site fermionic negativity
of $\rho$, which measures the $E_{ij}$ entanglement through the PPT criterion. 
As shown in Fig.\,\ref{fig:dimer}, 
the full fermionic negativity is generally larger 
than the QRE, except in some limiting cases. In fact it correctly 
captures the value of $2\log(2)$ in the noninteracting limit, 
\ie a maximally entangled dimer, leading to a maximally mixed single-site RDM.
In the strong-coupling limit ($U\gg t$), the negativity approaches the value 
of $\log(2)$, albeit significantly slower than the corresponding 
von Neumann entropy.

The overestimation of entanglement by negativities (dashed lines), 
with respect to geometric measures (solid lines), is observed also 
under symmetry resolution. 
In Fig.\,\ref{fig:dimer} we can see how all the three symmetry 
resolved terms are overestimated even in the noninteracting limit. 
While the antiferromagnetic and holon-doublon negativities approach
quite well the corresponding QRE measures in the strong coupling regime,
we observe that hopping negativity generally overestimates by a significant
amount the hopping entanglement, in the whole interaction range.

Notwithstanding these quantitative shortcomings, both the full and the
symmetry resolved negativities correctly capture the qualitative behavior
of the different entanglement terms. 
In particular, the holon-doublon and hopping terms are maximal
at the noninteracting point and decrease monotonically with an
increasing interaction strength, eventually vanishing in the 
strong-coupling limit. 
These features are well explained by two trivial observations. 
First, holon and doublon populations (given by $\mathcal{D} = N^2$)
are maximal at $U=0$ and decrease monotonically towards zero in the 
strong coupling limit. 
Furthermore, hopping processes are energetically
hindered by an increasing $U/t$ ratio.
On the other hand, the antiferromagnetic component increases
with the interaction strength $U/t$, saturating at a value of $\log(2)$ 
in the strong coupling limit ($U\gg t$), in good agreement with the 
evolution of the ground state Eq.\,\ref{eq:dimer_gs} towards a spin-singlet 
$\mid\!\!\text{GS}(U\gg t) \rangle \to \frac{1}{\sqrt{2}} \Bigl(\mid\uparrow\rangle \otimes \mid\downarrow\rangle -
\mid\downarrow\rangle \otimes \mid\uparrow\rangle \Bigr)$.
We note that the exact atomic limit ($t=0$) is 
well-known to be unentangled \cite{DissotiationParadoxSchilling}, 
as the ground state of the model becomes degenerate with the spin-triplet excited
states, and the inevitable mixture of these four states deletes all the 
antiferromagnetic entanglement, which is the only symmetry-resolved component
to approach a finite value in the $U\gg t$ regime.

\subsubsection{Effects of mixing in a noninteracting lattice}

From the exact solution of the isolated (pure) Hubbard dimer,
one might conclude that both the hopping and the holon-doublon
contributions to the two-site entanglement carry trivial insight
into Hubbard physics. Indeed, they appear to just describe a
crossover from a state featuring an equal population of empty, 
singly and doubly occupied sites, to the strong coupling regime,
where charge fluctuations are suppressed and the ground state is
dominated by spin-singlet configurations.
On the contrary, we will show that embedding the dimer into a 
lattice dramatically changes this picture, by the mere mixed 
character of the two-site reduced density matrix: in this setup,
the two-site RDM describes an open quantum system, with electrons
hopping from the dimer to the rest of the lattice and back.

In the noninteracting limit, the same Hamiltonian leads to the 
same equipartite population of the local Fock space. Yet, no 
holon-doublon or spin-singlet entanglement is present, as we
will readily show, for they are fully suppressed by the classical
mixing induced by the embedding into the lattice.
To this end, we evaluate the three symmetry-resolved components 
of the two-site entanglement for a reduced density matrix
$\rho^0_{\!\expval{ij}}$ describing two neighboring sites embedded
in a (noninteracting) tight-binding square lattice, at zero temperature. 
In the infinite lattice limit, one can prove that the one-body correlation matrix 
$\gamma^0 = \{\langle c^{\dagger}_{i,\sigma} c_{j,\sigma'} \rangle\}$ is given as
\cite{Bippus_Two-Site_Negativity_Hubbard}
\begin{equation}
 \langle c^{\dagger}_{i,\sigma} c_{j,\sigma'} \rangle  = 
    \begin{cases}
        \frac{\delta_{\sigma\sigma'}}{2} \textnormal{ \,, \, if \, }   (i_x,i_y) = (j_x,j_y)   \\[1ex]
        0  \textnormal{ \,, \, if \, }   (i_x-j_x)^2-(i_y-j_y)^2=0   \\[1ex]
        \displaystyle{-\frac{(-1)^{i_x-j_x}-(-1)^{i_y-j_y}}{\pi^2 \left[ (i_x-j_x)^2 - (i_y-j_y)^2 \right]}}\delta_{\sigma\sigma'}  \textnormal{, otherwise}   \\
  \end{cases}\;
\label{eq:noninteracting_square}
\end{equation}
where $i_x$, $i_y$, $j_x$, $j_y$ are the Cartesian coordinates for sites $i$ and $j$.

For nearest neighbors $\expval{ij}$ one can then restrict to
\begin{equation}
    \gamma^0_{\!\expval{ij}} = 
    \begin{pmatrix}
\frac{1}{2} & \frac{2}{\pi^2} & 0 & 0 \\[10pt]
\frac{2}{\pi^2} & \frac{1}{2} & 0 & 0 \\[10pt]
0 & 0 & \frac{1}{2}& \frac{2}{\pi^2} \\[10pt]
0 & 0 & \frac{2}{\pi^2} & \frac{1}{2}
\end{pmatrix},\label{eq:free_gamma}
\end{equation}
and compute $\rho_{\!\expval{ij}}$ with Peschel's formalism \cite{Peshel_2001,Peschel2003,Peschel_2009}, 
as for noninteracting systems any many-body expectation value is fully determined by one-body correlation
matrix elements via Wick contractions \cite{GaussianRMP,Adesso2014_Gaussian,CostaDias2022_GaussianPartialTraces}.
In particular, one can prove that $\rho^0_{\!\expval{ij}}$ can be written as the Gaussian density matrix
\begin{equation}
    \rho^0_{\!\expval{ij}} = K \exp\left(-\sum_{\expval{ij}} \sum_{\sigma\sigma'} H^\mathrm{e}_{i,\sigma j,\sigma'}c^\dagger_{i,\sigma}c_{j,\sigma'}\right),
\end{equation}
where $K$ is a normalization factor and 
$H^\mathrm{e}$ is the quadratic \emph{entanglement Hamiltonian},
determined by the one-body correlation matrix as
\cite{Peshel_2001,Peschel2003,Peschel_2009}
\begin{equation}
    (H^\mathrm{e}_{\!\expval{ij}})^\intercal= \log(\frac{\mathbf{1}-\gamma^0_{\!\expval{ij}}}{\gamma^0_{\!\expval{ij}}}).
    \label{eq:peschel}
\end{equation}

From $\rho^0_{\!\expval{ij}}$ we can then get exact values of the full and symmetry resolved entanglement measures,
for the considered nearest-neighbor dimer in the noninteracting square lattice. We obtain
\begin{equation}
    \mathcal{N}_{ij}^{\uparrow\downarrow} = \mathcal{N}_{ij}^\mathrm{hd} = 0, \quad
    \mathcal{N}_{ij}^{\mathrm{F}t} \simeq 0.44 \log(2), \quad  
    I_{ij} \simeq 0.49 \log(2), \quad S_{ij} \simeq 3.51 \log(2) 
\end{equation}
Remarkably, the two superselected entanglement components are exactly zero, 
as guaranteed by the PPT criterion, which is faithful for the small 
antiferromagnetic and holon-doublon sectors. 
The only surviving term is the hopping one, which for a mixed state 
we can estimate only through the $\mathcal{N}_{ij}^{\mathrm{F}t}$ negativity.
Though finite, the hopping negativity is significantly reduced with respect to
the noninteracting value in the pure Hubbard dimer ($\simeq 1.59 \log(2)$).  
The mutual information, bounding from above both entanglement and classical 
correlations, is much smaller than the $4\log(2)$ value realized in the 
noninteracting pure Hubbard dimer. This reduction is directly related 
to the growth of the two-site von Neumann entropy from zero to $3.51 \log(2)$.
In fact, the mixed character of the two-site reduced density matrix, 
induced by the embedding in a highly delocalized (metallic) lattice 
and quantified by a large value of $S_{ij}$ accounts for a drastic 
reduction of entanglement between the two sites, despite the system
being described by the same formal Hamiltonian. 

Similar results, at finite interaction strength $U/t$,
have been found for thermal or dephasing-induced 
mixing of the two-site density matrix, in
analytical treatments of the Hubbard dimer
\cite{Dowling2006,thermal_concurrence_Hubbard_dimer,thermal_spin/charge_concurrence_Hubbard_dimer,Zavatti_Magic_Hubbard_dimer,Canio_concurrence_Hubbard_dimer} 
and the single-bath-site
Anderson model \cite{Canio_orbital_concurrence}, as well as
in numerically exact simulations of small Hubbard clusters
\cite{thermal_death_ssr}.
In Section \ref{sec:cdmft_results} and in Appendix 
\ref{app:risb}, we discuss the effect of 
embedding-induced mixing at finite $U/t$, 
considering numerical data from quantum cluster
methods, across a wide parameter range.

\section{Quasilocal entanglement in the Hubbard model: plaquette CDMFT}
\label{sec:cdmft_results}


In the following we present cellular dynamical mean-field theory (CDMFT) 
\cite{CDMFT_ORIGINAL,Kotliar2001,Biroli2002} simulations of the two-dimensional 
Hubbard model on the square lattice, at zero temperature. 
With respect to the standard dynamical mean-field theory \cite{DMFT_RMP}, 
CDMFT embeds a finite bounded cluster in the auxiliary self-consistent bath. 
In this way, on top of the exact treatment of the local dynamical self-energy,
it injects numerically exact short-range correlations, defined within the given
cluster size. 

As the cluster problem is solved with open boundary conditions, the
periodicity of the original unbounded lattice is partially broken. 
Different, inequivalent periodization schemes have been proposed 
\cite{Senechal_gscheme,Capone2004_1dCDMFT,Sangio_CDMFTsizescaling,center-focused-CDMFT,Schafer24Mott,Tremblay_compact_tiling,Pelz_LiouvillianCDMFT},
with no established consensus on subtle properties in momentum space,
such as the Fermi-arc versus hole-pocket nature of the pseudogapped 
Fermi surface in the underdoped Hubbard model
\cite{Capone_Pseudogap_CDMFTanisotropy,Kotliar_Pseudogap_CDMFT_AFMcorr,Kotliar_Pseudogap_CDMFTzerosGF,Imada_Pseudogap_CDMFT}.
Nevertheless, real-space quantities defined within the cluster are naturally 
represented in the method.

In this work, we carry out all calculations with a $2\times2$
cluster (``\emph{plaquette}''), so that the two independent two-site reduced 
density matrices $\rho_{\!\expval{ij}}$ and $\rho_{\!\expval{\expval{ij}}}$, 
respectively for nearest and next-nearest neighbors, are well defined quantities.
To evaluate them we perform direct tracing of the bath and cluster degrees of freedom 
in the ground state of the auxiliary cluster impurity model \cite{BellomiaPhD}.
More details on the Hamiltonian parametrization of the bath and our implementation
of Krylov-based exact diagonalization, based on the \textsc{EDIpack} library 
\cite{EDIpack,EDIpack2,EDIpack_code}, can be found in Appendix \ref{app:cdmft_ed}.

Given the size of the addressed cluster, our calculations are meant to address 
only the \emph{quasilocal} component of the two-site entanglement.

\subsection{Interaction-driven MIT at half-filling} 
\label{sec:half_filled_MIT}

To begin with, we revisit the paramagnetic metal-insulator transition (MIT)
in the half-filled two-dimensional Hubbard model (Eq.~\ref{eq:hubbard}). 
At large electronic repulsion, the paramagnetic solution of the model
describes a Mott insulator characterized by the presence of incoherent 
high-energy excitations and the absence of low-energy quasiparticles 
\cite{Mott_RMP,MIT_Bethe_1992,Gebhard1997,Balzer2009_MottVCA,MitchellTopoMott,D-TRILEX_2dHubbard}. 
This state, universally considered a cornerstone of strongly correlated 
electrons, has recently been the subject of intense research efforts, 
aimed at characterizing the strong correlation effects in terms of 
quantum information concepts
\cite{Vollhardt_nonfreeness,XiDai2013,Cocchi_ColdAtoms,WalshLetter2019,Bellomia2023,bellomia_local_classical,more_is_uncorrelated}.
In this context, in a previous work about the short-range entanglement across
the interaction-driven MIT, we have pinpointed the charge superselected 
entanglement as a defining property of the paramagnetic Mott-Hubbard
insulator in CDMFT, for its value does not change with increasing interaction, 
within the Mott localized state \cite{Bellomia2023}. 
Yet, the behavior of the full entanglement across the transition 
and its symmetry resolution were 
not explicitly addressed, except for the upper bound given
by the two-site mutual information, which includes also classical correlations
\cite{Vedral2001,Discord&Classical_unified}.

Here we extend the analysis to characterize the full two-site entanglement
in the half-filled MIT through the fermionic negativity and, more 
importantly, we analyze its symmetry-resolution in spin-singlet,
holon-doublon and hopping-induced contributions.
As thoroughly discussed in Section \ref{sec:sym-res}, the first two terms
are directly connected to the superselected entanglement studied in 
\cite{Bellomia2023} through closed formulas for the REE.  In the following
we report also the corresponding antiferromagnetic and holon-doublon REE, 
for a more direct comparison. Additionally,
we recall that entanglement measures defined through the quantum relative
entropy provide a more accurate quantification of the entanglement encoded
in the partial transpose, with respect to the fermionic negativity. In fact,
the REE evaluates the quantum distance between the given density matrix (or
a given symmetry sector therein) and its closest separable state.

\begin{figure*}
    \centering
    \includegraphics[width=\linewidth]{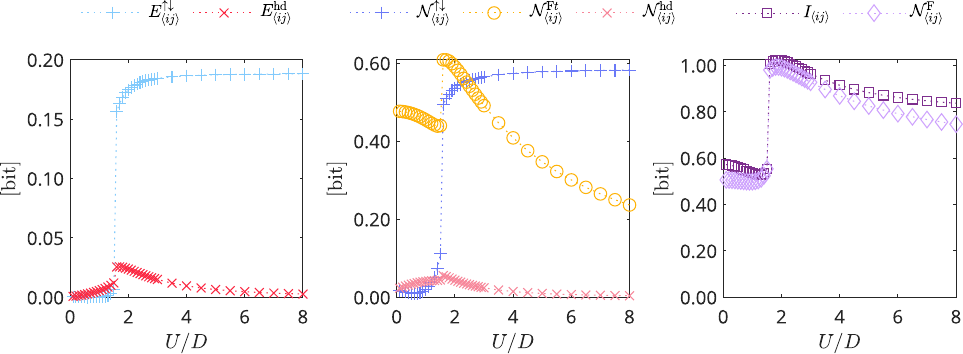}
    \caption{Behavior of symmetry-resolved REE (left), 
    symmetry-resolved negativity (middle)
    and measures of full entanglement (right) 
    between two neighboring sites, across the 
    interaction-driven Mott transition.}
    \label{fig:half_filling}
\end{figure*}

\smallskip 
In Fig.\,\ref{fig:half_filling} we show numerical results for the superselected 
symmetry components of the REE (left panel), the full symmetry resolution of the
fermionic negativity (middle panel) and the two measures of full entanglement 
and correlations (right panel), for the nearest-neighbors density matrix
$\rho_{\!\expval{ij}}$ extracted from our $2\times2$ CDMFT calculations.
We note that in \cite{Bellomia2023} larger clusters were studied, providing strong
evidence for an extremely fast decay of all correlations with distance, across the
whole interaction range. For this reason we will not explicitly discuss here results
from the next-nearest-neighbors density matrix $\rho_{\!\expval{\expval{ij}}}$, 
other than noting that the antiferromagnetic contribution to the entanglement
vanishes for all computed points, while the holon-doublon one is finite only
well inside the Mott phase ($U\gtrsim 2D$) and remains tiny, with
$\mathcal{N}_{\!\expval{\expval{ij}}}^{\mathrm{hd}}\lesssim 5\times10^{-4}$
bit and
$E_{\!\expval{\expval{ij}}}^{\mathrm{hd}}\lesssim 10^{-4}$ bit. 
Hence, the full two-site entanglement in $\rho_{\!\expval{\expval{ij}}}$ is dominated by 1-charge jump amplitudes, with $\mathcal{N}_{\!\expval{\expval{ij}}}
^{\mathrm{F}}\lesssim 10^{-2}$ bit, and the absence of a next-nearest-neighbor 
hopping in the Hamiltonian provides a clear reason for the significant damping 
with respect to nearest neighbors. A more detailed analysis of the spatial structure of the full two-site negativity in the weak-coupling regime of the
model has been provided by quantum Monte Carlo simulations and the parquet 
dynamical vertex approximation \cite{Bippus_Two-Site_Negativity_Hubbard}.

\subsubsection{Spin-singlet entanglement as a robust feature of the Mott phase}

In the weak coupling regime, the nearest-neighbor antiferromagnetic REE
$E_{\!\expval{ij}}^{\uparrow\downarrow}$ is suppressed because of the excessive
mixing between spin-singlet and spin-triplet states.
Crossing the Mott transition (located at $U\simeq1.55 D$ in our CDMFT
calculation, where $D=4t$ is the half-bandwidth of the noninteracting 
lattice dispersion) its value is however greatly enhanced:
after a quick transient, it saturates to 
$E_{\!\expval{ij}}^{\uparrow\downarrow}\approx 0.19$ bit 
and remains stable against variations of the $U/D$ ratio. This suggests 
that the formation of quasilocal spin singlets is a characteristic feature
of the Mott phase, not particularly sensitive to details and parameters in
the Hamiltonian. The corresponding antiferromagnetic negativity 
$\mathcal{N}_{\!\expval{ij}}^{\uparrow\downarrow}$ shares the same essential
behavior, modulo an overestimation of the saturation value in the $U\gg D$ regime,
as for the exact ground state of the Hubbard dimer (see Section \ref{sec:dimer}).

\subsubsection{Role of holon-doublon entanglement in driving the MIT}

As for the antiferromagnetic component, also
the holon-doublon entanglement $E_{\!\expval{ij}}^\mathrm{hd}$ vanishes
in the noninteracting limit, due to mixing.  Yet, it increases slightly
faster with the $U/D$ ratio, with respect to spin-singlet entanglement. 
At the Mott transition the value of $E_{\!\expval{ij}}^\mathrm{hd}$ is 
also bumped, but much less dramatically than for the antiferromagnetic component.
Finally, inside the Mott insulator, the holon-doublon entanglement is 
progressively damped by the increasing interaction strength, as there
are fewer and fewer empty and doubly-occupied sites to be bound. 
In fact, for $U\gg D$ we observe $E_{\!\expval{ij}}^\mathrm{hd} \to 0$.

Overall, the holon-doublon entanglement is appreciable only near the MIT 
-- more so on the insulating side -- reconnecting to early proposals of 
nearest-neighbor binding of holons and doublons as a key mechanism driving
Mott localization 
\cite{Watanabe2006_holon_doublon_binding,Zhou2014_holon_doublon_binding,Prelov2015_holon_doublon_binding}. 
\gb{Indeed, as shown in Section \ref{sec:sym-res-mixed}, a finite 
holon-doublon entanglement requires nearest-neighbor holon-doublon
configurations to be favored over holon-holon and doublon-doublon
ones, as well as coherently superposed.}
Here 
we note that an account of the holon-doublon binding has long been 
deemed inaccessible to dynamical mean-field theory and its cluster 
extensions \cite{Capello_VMC_Mott}. 
We overcome this issue via the symmetry resolution of the superselected entanglement.

As for the spin-singlet component, also the holon-doublon entanglement is well
captured by the correspondent negativity $\mathcal{N}_{\!\expval{ij}}^\mathrm{hd}$,
at least as an upper bound. We observe that the two superselected negativities do
not vanish in the $U\ll D$ regime. In Appendix \ref{app:risb}, we elaborate on why 
this is an artifact of the CDMFT calculation, and is expected to be solved by 
employing a larger auxiliary bath 
or alternative cluster embedding approaches that directly optimize static quantities.
On the other hand, results at strong coupling and near the MIT are reliable.

\subsubsection{Beyond superselection rules: hopping and full entanglement}

Looking past the superselection rules, whose role is to single out usable quantum
resources, we can determine the entanglement generated by hopping processes. 
As discussed in Section \ref{sec:sym-res-mixed}, for mixed states this term 
cannot be evaluated in the REE framework. Thus we rely on the hopping
negativity $\mathcal{N}_{\!\expval{ij}}^{\mathrm{F}t}$, as defined in Section
\ref{sec:sym-res-mixed}. 
Its behavior, reported in the central panel of Fig.\,\ref{fig:half_filling},
generally reflects the competition between the two relevant energy scales 
in the problem, the interaction $U$ and the half-bandwidth $D=4t$. 
As we increase the $U/D$ ratio, the hopping entanglement gets monotonically 
reduced, on either side of the MIT. However, at the MIT the overall scale of
this term is significantly increased, confirming that all kinds of quantum 
correlations are enhanced when entering the Mott regime. 

In fact, the full negativity $\mathcal{N}_{\!\expval{ij}}^{\mathrm{F}}$ 
shares the same general features, and so does
also the mutual information $I_{\!\expval{ij}}$
(right panel in Fig.\,\ref{fig:half_filling}).
This indicates that also classical correlations may increase at the MIT.
The close similarity -- also on a quantitative level -- between the two-site 
negativity and the two-site mutual information suggests that, despite 
the absence of a rigorous reverse PPT criterion for the hopping sector
in the partial transpose, we are not severely underestimating the possible
nondistillable entanglement, as the associated correlations are also 
bounded from above by the mutual information.

Finally, we note that the large-$U$ evolution of the full quantities is 
significantly slower than what is observed for $\mathcal{N}_{\!\expval{ij}}^{\mathrm{F}t}$.
This is simply a result of the nearly constant value of the spin-singlet
entanglement at strong coupling, which must be contained in both 
$\mathcal{N}_{\!\expval{ij}}^{\mathrm{F}}$ and
$I_{\!\expval{ij}}$.

\subsubsection{Entanglement portrait of the MIT and comparison with the pure dimer}

Before moving to the hole-doped regime of the model, we comment on the
relative magnitude of the three different symmetry-resolved components of 
the nearest-neighbor entanglement, at half-filling. To do so we will focus
on the negativity measures (central panel in Fig.\,\ref{fig:half_filling}).
In the weak coupling regime ($U$ under the critical interaction for the MIT)
the hopping entanglement fully dominates the picture. At strong coupling, 
$U\gg D$, the dominant component is the spin-singlet entanglement, while the
other two are progressively damped. Close to the MIT all entanglement terms
are enhanced, but the holon-doublon component is markedly lower in
magnitude with respect to the other two, comparable, terms.

We end the section by noting that in the Mott phase all nearest-neighbor
quantities display the same essential behavior as in the (pure) ground 
state of the isolated dimer.
Even if all values are much reduced, proving that mixing is still present
and significant, this observation is in good agreement with the expectation
that the Mott state is much more localized than the itinerant Fermi liquid. 

In the next section we will
discuss how the itinerant metal found at low hole doping is much closer to 
the entanglement properties of the Mott insulator -- hence of an isolated 
Hubbard dimer -- than those of the Fermi liquid found at large hole doping.

\subsection{Doping-driven MIT and pseudogap physics} 
\label{sec:doped_MIT}

In this section we proceed to dope the Mott phase found in the 
CDMFT calculation discussed in the previous Section
\ref{sec:half_filled_MIT}. Among all possible choices for the
interaction strength we take $U=2.3D$. As it is not too far into the 
localized phase, we have both an appreciable holon-doublon entanglement 
and similar magnitudes for the spin-singlet and hopping negativities. 

\begin{figure}
    \centering
    \includegraphics[width=0.7\linewidth]{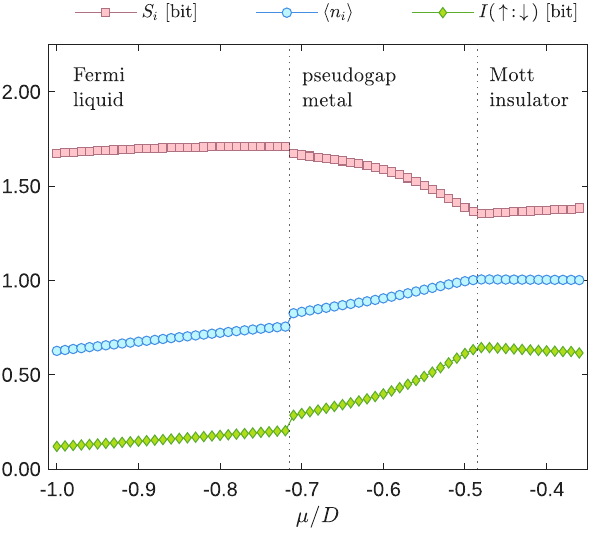}
    \caption{Local von Neumann entropy $S_i$\cite{Franca_LDAentropy,XiDai2013,Cocchi_ColdAtoms,WalshLetter2019,WalshPRXQuantum2020,Franca_local_entropy_superc},
    local electron density $\langle n_i \rangle$ and
    local nonfreeness $I(\uparrow:\downarrow)$\cite{Gottlieb2005,Gottlieb2007,Vollhardt_nonfreeness,Held_nonfreness,bellomia_local_classical,more_is_uncorrelated}
    across the density-driven Mott transition.
    All data points pinned at $\langle n_i \rangle=1$ (half-filling),
    represent the incompressible Mott insulator. At intermediate 
    density $0.83\lesssim\langle n_i \rangle<1$, the system is delocalized 
    into a \emph{strongly correlated} -- meaning large $I(\uparrow:\downarrow)$ --
    metal, akin to the pseudogap regime of cuprates. 
    Finally, the metal found for $\langle n_i \rangle \lesssim 0.76$
    features a full Fermi surface and becomes \emph{weakly correlated}
    -- meaning vanishing $I(\uparrow:\downarrow)$ -- at very low electron density. 
    }
    \label{fig:_locent_njump_nongauss}
\end{figure}

To inject holes in our calculations, we gradually decrease the
chemical potential $\mu$ in the auxiliary cluster Hamiltonian 
(see Eq.\,\ref{eq:H_cluster} in Appendix \ref{app:cdmft_ed}).
As the Mott insulator is incompressible, there are many values of the
chemical potential for which the CDMFT solution does not change,
apart from some limited numerical drift. Once the electron density 
starts to change, we observe a progressive delocalization of electrons,
as can be inferred from the evolution of the single-site von Neumann
entropy $S_i$, which provides an upper bound for local charge fluctuations
\cite{BellomiaPhD}.
We show the $\mu$-dependence of $S_i$ in Fig.\,\ref{fig:_locent_njump_nongauss}
(pink squares) and refer the reader to \cite{WalshPRXQuantum2020} for a thorough 
analysis of its behavior in the ($U$, $T$, $\mu$) parameter space.
At larger doping, the system undergoes a discontinuous jump in all
observables, \eg the electron density (see light blue circles in
Fig.\,\ref{fig:_locent_njump_nongauss}), marking a first-order 
transition to a different kind of metal. 
This phenomenology, overall, describes a two-step Mott
transition, as we discuss below.

At small $|\mu|/D$, one has a \emph{crossover} between the incompressible
insulator -- pinned at half-filling -- and a strongly correlated metal,
which features a moderate hole density ($\delta = 1-n \lesssim 0.2$) and shares 
many spectral properties with the pseudogap regime of cuprate superconductors 
\cite{Capone_Pseudogap_CDMFTanisotropy,Kotliar_Pseudogap_CDMFT_AFMcorr,Kotliar_Pseudogap_CDMFTzerosGF,Imada_Pseudogap_CDMFT}. We quantify the correlated character of
this regime with the local nonfreeness $I(\uparrow:\downarrow)$ 
\cite{bellomia_local_classical}, which measures inter-particle correlations 
as the distance from the set of free-electron states 
\cite{Gottlieb2005,Gottlieb2007,Vollhardt_nonfreeness,Held_nonfreness,more_is_uncorrelated}.
We report its $\mu$-dependence in Fig.\,\ref{fig:_locent_njump_nongauss} 
(green diamonds).

At higher doping fractions, this pseudogap-like state undergoes an abrupt 
transformation into a less correlated metal, featuring a full Fermi surface
(a standard Fermi liquid).
The values of the local nonfreeness in Fig.\,\ref{fig:_locent_njump_nongauss}
attest that the Fermi liquid still holds a significant degree of electronic
correlation in proximity of the first-order transition. The system eventually
becomes weakly correlated at extreme doping fractions ($\delta\simeq0.4$).

This scenario has been carefully analyzed in the past, within 
analogous plaquette CDMFT calculations at finite (including low) temperature 
\cite{Sordi_njumps_PRL,Sordi_njumps_PRB}, as well as in larger periodic
clusters for the triangular lattice \cite{Gingras_njumps_DCA}. 
Of course, some specific details of the physical picture, 
such as the order of the metal-metal transition, can change 
on the square lattice by considering larger clusters or
clever momentum patching
\cite{Jarrell_Pseudogap_DCA_AFMcorr,Gull_Pseudogap_DCA_k-selective,Gull_Pseudogap_DCA_k-patches,Ferrero_PG_as_k-selective_Mott,Ferrero_Pseudogap_Lifshitz,Toschi_Pseudogap_spinfluct,Krien_corr_self-energy,Pelz_criticality_pseudogap_letter,Pelz_criticality_pseudogap_longer},
or -- more in general -- long range correlations beyond cluster theories
\cite{RenyiPseudogap}. Nonetheless, our analysis is 
independent of such details, as it characterizes the pseudogap
metal as having \emph{qualitatively} different entanglement properties,
under symmetry resolution, with respect to the overdoped Fermi liquid. 

Remarkably, a similar qualitative distinction has recently been observed 
in a quantum gas simulator, where the spin-singlet contribution to the 
superselected entanglement has been measured at the high-$T$ onset of the
pseudogap regime of a large Fermi-Hubbard system, while entanglement
beyond the antiferromagnetic sector has been investigated with the 
lambda-corrected dynamical vertex approximation \cite{SSR_PG_ColdAtoms}.
We note that the spin-singlet entanglement is confined to nearest
neighbors in the quantum gas simulation of the model, and the 
remaining entanglement decays very fast with distance in the dynamical 
vertex calculations. Hence, as for the half-filled MIT, our analysis 
will be focused on the nearest-neighbor density matrix only.

\begin{figure*}[t!]
    \centering
    \includegraphics[width=\linewidth]{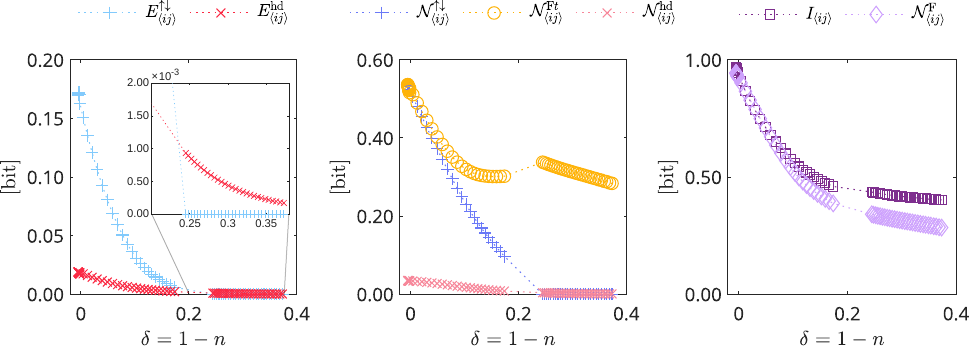}
    \caption{Behavior of symmetry-resolved REE (left), symmetry-resolved negativity (middle)
    and measures of full entanglement (right) between two neighboring sites, across the 
    density-driven Mott transition. The inset highlights how the antiferromagnetic entanglement
    is completely absent in the overdoped metal, while the holon-doublon term is small 
    but finite at all the considered doping fractions.}
    \label{fig:hole_doping}
\end{figure*}

\smallskip

In Fig.\,\ref{fig:hole_doping} we present our results 
for the nearest-neighbor measures of full and symmetry resolved entanglement,
as a function of the hole-doping fraction $\delta=1-n$. 
The range of doping with no data points marks the first-order transition 
between the underdoped and the overdoped metals, namely the discontinuous
jump in the local electron density (see light blue circle in 
Fig.\,\ref{fig:_locent_njump_nongauss}).
The cluster of points at $\delta=0$ corresponds to calculations for different
values of the chemical potential, but all confined in the Mott 
insulating phase.


\subsubsection{Failure of the full two-site entanglement}

To begin with, we note how the full negativity and the mutual information (right panel in Fig.~\ref{fig:hole_doping})
correctly account for the general weakening of correlations with doping,
similarly to the local nonfreeness $I(\uparrow:\downarrow)$ (green diamonds
in Fig.\,\ref{fig:_locent_njump_nongauss}).
Yet, these quantities completely fail at discriminating between the three 
different regimes.
In particular, excluding the discontinuous jump appearing in any quantity, 
which of course locates the metal-metal transition, there are no genuinely 
different features distinguishing the two metals at its left and right.
Moreover, it is not possible to infer a fundamentally different degree of
correlation or entanglement between the pseudogap metal at very low $\delta$
and the half-filled insulator.
In the nearest-neighbor negativity and mutual information, the density-driven
MIT appears as a simple gradual crossover between a strongly correlated regime
at low doping and a weakly correlated regime at very large doping.
On the contrary, as we will discuss in the following, the symmetry resolution
of the quasilocal entanglement provides a sensible characterization of the solutions. 

\subsubsection{Spin-singlet entanglement spills out of half-filling, into the pseudogap}

By looking only at the antiferromagnetic contribution to the
nearest-neighbor entanglement, measured by either 
$E_{\!\expval{ij}}^{\uparrow\downarrow}$ or
$\mathcal{N}_{\!\expval{ij}}^{\uparrow\downarrow}$, we unveil a distinctive
property of the pseudogap state. Although metallic, hence featuring 
delocalized electrons at least in some regions of momentum space
\cite{Capone_Pseudogap_CDMFTanisotropy,Kotliar_Pseudogap_CDMFT_AFMcorr,Kotliar_Pseudogap_CDMFTzerosGF,Imada_Pseudogap_CDMFT},
this regime preserves the localized spin-singlets that in 
Section \ref{sec:half_filled_MIT} we have identified as a robust property 
of the Mott insulating phase. 
Remarkably, as for the weakly-interacting Fermi liquid (see 
Fig.\,\ref{fig:half_filling} and also Fig.\,\ref{fig:CRISBvCDMFT} in 
Appendix \ref{app:risb}, for a more detailed analysis), in the 
overdoped metal at $U=2.3D$, the spin-singlet component of the entanglement
exactly vanishes. Thus, according to the PPT criterion, the antiferromagnetic
sector in $\rho_{\!\expval{ij}}$ is separable.
It follows that the partial delocalization of the injected hole carriers in
the underdoped metal does not mix spin-singlet and spin-triplet components 
to a sufficient degree as to kill the spin entanglement in the two-site subsystem. 
This puts the CDMFT description of the pseudogap metal in the same 
spin-entanglement class of Mott insulators, markedly distinct from standard Fermi
liquid metals, regardless of the weakly or strongly interacting regime in which the latter are found.
We remark once again that, at high temperatures close to where the
pseudogap is closed by thermal fluctuations, a similar clear-cut connection 
between the pseudogap and the nearest-neighbor spin-singlet entanglement has
been found also in large Hubbard systems, simulated with either diagrammatic
extensions of DMFT or cold-atom experiments \cite{SSR_PG_ColdAtoms}.

We highlight that these observations are in good agreement with the widespread
idea that pseudogap physics is tightly connected with antiferromagnetic
fluctuations and correlations
\cite{Jarrell_Pseudogap_DCA_AFMcorr,Kotliar_Pseudogap_CDMFT_AFMcorr,Toschi_Pseudogap_spinfluct,Krien_corr_self-energy,Ferrero_Pseudogap_Lifshitz,Schafer24Mott,RenyiPseudogap}.
Yet, the usual diagnostics have historically neglected the classification
in classical and quantum character of these spin correlations. The behavior
of the spin-singlet entanglement shows that the truly quantum part of the 
antiferromagnetic origin of pseudogap physics appears to be extremely well
localized and smoothly connected to the properties of the Mott insulator, 
while absent in the overdoped Fermi liquid regime.

\subsubsection{Holon-doublon entanglement survives the closure of the pseudogap}

Moving forward to the other contribution to the superselected entanglement,
the holon-doublon term, we observe that it behaves quite similarly to
antiferromagnetic entanglement, as long as the crossover between the Mott
insulator and the pseudogap regime is concerned. 
Yet, at the first-order transition towards the Fermi liquid metal the 
holon-doublon entanglement shows an important qualitative difference.
While the antiferromagnetic sector is exactly separable, owing
to the PPT criterion, the holon-doublon sector remains entangled, even
though the residual values are quantitatively small
(see inset in Fig.\,\ref{fig:hole_doping}). 

\begin{figure*}[ht!]
    \centering
    \includegraphics[width=.9\linewidth]{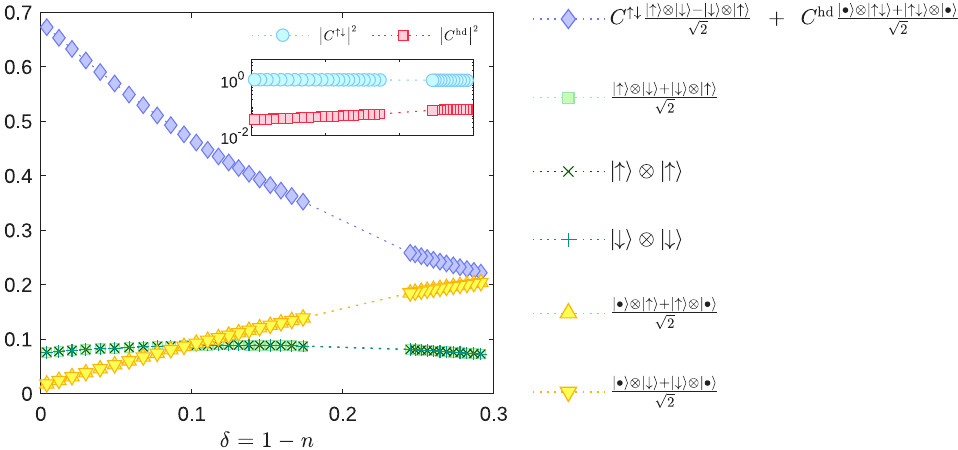}
    \caption{Doping evolution of the dominant eigenstates in the nearest-neighbor RDM. 
    Each plotted eigenvalue determines the corresponding classical weight in 
    $\rho_{\!\expval{ij}}$. Triplet states are degenerate because of SU(2) symmetry. 
    The holon-spin states are also invariant under spin-flip transformations. 
    The inset shows the quantum probabilities associated to the spin-singlet and 
    holon-doublon Bell pairs that contribute to the first dominant eigenstate. 
    The parity superselection rule splits this quantum superposition into a classical
    mixture of the two, leading to the symmetry resolution into antiferromagnetic and 
    holon-doublon entanglement (Fig.\,\ref{fig:hole_doping}).}
    \label{fig:dominant_states}
\end{figure*}

A better understanding of this behavior is obtained analyzing the
spectral decomposition of $\rho_{\!\expval{ij}}$. 
In Fig.\,\ref{fig:dominant_states} we show the six largest eigenvalues
of the nearest-neighbor density matrix.
Each of these weights is associated to a two-site pure state, bringing 
its own entanglement properties in the mixture defining $\rho_{\!\expval{ij}}$.
We recall that, crucially, as discussed in Section \ref{sec:dimer}, the 
classical mixing of different pure states can make quantum amplitudes 
cancel out, and so dampen or eventually kill entanglement.
For all the relevant doping fractions, the largest weight 
$\lambda_{\text{dominant}}$ in $\rho_{\!\expval{ij}}$ is associated to
a pure state in the form
\begin{equation}
    \ket{\text{dominant}} = 
    C^{\uparrow\downarrow} \frac{\ket{\uparrow}\otimes\ket{\downarrow}
    -\ket{\downarrow}\otimes\ket{\uparrow}}{\sqrt{2}} +
    C^\mathrm{hd} \frac{\ket{\bullet}\otimes\ket{\uparrow\downarrow}
    +\ket{\uparrow\downarrow}\otimes\ket{\bullet}}{\sqrt{2}}.
\end{equation}
This state clearly violates the parity superselection rule, as it describes
a quantum superposition of a spin-singlet (odd number of electrons on 
each single site) and a holon-doublon pair (even number of electrons).
After applying the P-SSR, this state is split into a classical mixture of
the spin-singlet and the holon-doublon Bell pairs, whose relative weights
are given by $|C^{\uparrow\downarrow}|^2$ and $|C^\mathrm{hd}|^2$. These
two weights, combined with $\lambda_{\text{dominant}}$, determine the 
entanglement content in the spin-singlet and holon-doublon sectors. 
As the dominant eigenvalue is monotonically decreasing with $\delta$, it
is not surprising that both $E_{\!\expval{ij}}^{\uparrow\downarrow}$
and $E_{\!\expval{ij}}^\mathrm{hd}$ similarly decrease by injecting holes
in the system. In fact, the decreasing of the average electron density 
naturally dampens the number of doublons available to entangle with holons. 
The resulting delocalization of carriers out of the Mott blockade does in 
turn enhance the mixing of singlet and triplet states in the 
antiferromagnetic sector, decreasing $E_{\!\expval{ij}}^{\uparrow\downarrow}$
until it eventually dies. The three degenerate spin-triplet eigenvalues 
(see green markers in Fig.\,\ref{fig:dominant_states}) reach the critical 
value of $1/3$ of the spin-singlet weight at the metal-metal transition.

\gb{The different fate of the two superselected components can be traced back
to the symmetries of the diagonal populations entering the PPT conditions
(Eqs.\,\ref{eq:nssr_condition} and \ref{eq:pssr_condition}).
In the antiferromagnetic sector, the $\mathrm{SU}(2)$ symmetry enforces
$\rho_{\!\expval{ij}}(4,4) = \rho_{\!\expval{ij}}(13,13)$, so that the
geometric mean in Eq.\,\ref{eq:nssr_condition} coincides with the population
of each parallel-spin configuration.
The spin-singlet amplitude must then overcome the full weight
of these configurations, and the antiferromagnetic entanglement strictly
vanishes as soon as the spin-triplet eigenvalues reach the critical threshold.
In the holon-doublon sector, the analogous role is played by particle-hole
symmetry, which at half filling enforces
$\rho_{\!\expval{ij}}(1,1) = \rho_{\!\expval{ij}}(16,16)$.
Hole doping, however, breaks this symmetry more and more: the holon-holon
population $\rho_{\!\expval{ij}}(1,1)$ grows, while the doublon-doublon
population $\rho_{\!\expval{ij}}(16,16)$ is strongly damped, as pairs of
neighboring doublons -- already suppressed by the Hubbard repulsion --
become extremely rare. Hence, the geometric mean
$\sqrt{\rho_{\!\expval{ij}}(1,1)\,\rho_{\!\expval{ij}}(16,16)}$ remains much
smaller than the larger of the two populations, keeping the threshold in
Eq.\,\ref{eq:pssr_condition} low.
At the same time, the dominant eigenstate contributes to the holon-doublon
amplitude $\rho_{\!\expval{ij}}(6,11)$ with a term
$\lambda_\text{dominant}|C^\mathrm{hd}|^2/2$.
While $\lambda_\text{dominant}$ decreases by a factor of three across the
considered doping range, the relative weight $|C^\mathrm{hd}|^2$ of the
holon-doublon pair within the dominant state increases from about $0.04$
at half filling to about $0.06$ at the metal-metal transition, and up to
about $0.09$ in the overdoped metal (see inset in Fig.\,\ref{fig:dominant_states}),
so that this contribution is reduced by only about 20\%.
As a result, while the antiferromagnetic entanglement dies at the metal-metal
transition, the holon-doublon entanglement develops a flat, small but finite
tail in the overdoped metal. The increasing holon-holon population and the
reduced number of doublons shrink the negative eigenvalue of the holon-doublon
block in the partial transpose, but the vanishingly small doublon-doublon
population prevents the product of the diagonal terms from overcoming the
residual holon-doublon amplitude.}

\subsubsection{Dominant contribution to the hopping entanglement}

To end the section, we discuss briefly some features of the hopping-induced 
negativity in Fig.\,\ref{fig:hole_doping}. Overall, the doping
evolution of this quantity in the pseudogap metal follows closely
the behavior of the full fermionic negativity. In fact, in this
regime, it is the dominant contribution to the full entanglement
between the nearest neighboring sites, together with the spin-singlet
contribution.
Yet, we observe that around 10\% doping, the two quantities significantly
diverge: while the antiferromagnetic negativity keeps decreasing steadily,
the hopping negativity slows down to a plateau value at the metal-metal
transition point. Notably, around the same doping threshold, the two
dominant contributions to the hopping negativity, namely the spin-holon
Bell pairs whose weights are depicted by yellow triangles in
Fig.\,\ref{fig:dominant_states}, surpass the weight of the spin triplet in
the whole $\rho_{\!\expval{ij}}$ density matrix. Hence, close to
the metal-metal transition, the pseudogap regime -- while still dominated
by spin-singlets -- is significantly composed of spin-holon pairs at the
quasilocal level. In fact, spin-holon pairs constitute the second dominant
contribution to the nearest-neighbor reduced density matrix. 

After the metal-metal transition, as discussed earlier, the antiferromagnetic
entanglement vanishes due to the mixing between spin-singlet and spin-triplet
states. On the contrary, the hopping entanglement, which is dominated by spin-holon pairs, is boosted and resumes
the general decreasing trend towards the trivial empty lattice. Hence, we 
remark that, while not having anymore the partial Mott-like character of the
pseudogap metal, the Fermi liquid after the discontinuous transition is still
far from trivial, in its entanglement properties.
This explains the reason for the failure of the full two-site entanglement
in distinguishing the two metals at the qualitative level.

We end the section by noting that the significant presence of spin-holon entangled
pairs, with a weight that increases with the doping fraction, recalls some 
characteristic features of a fractionalized SU(2) gauge theory for the pseudogap 
of cuprates \cite{Sachdev_Pseudogap_Toporder,Ferrero_Pseudogap_Lifshitz}.
Yet, we clarify that our representation assumes no fractionalization between 
charge and spin degrees of freedom in the quasilocal density matrix. 
Additionally, while the spin-holon contribution to $\rho_{\!\expval{ij}}$
as well as the overall hopping-induced entanglement are both boosted at the
metal-metal transition, the overall magnitude of the latter is indeed larger
close to half filling, than in the overdoped Fermi liquid. Finally, the 
hopping-induced negativity entails many other off-diagonal amplitudes than the 
simple spin-holon pairs discussed above. Even if the spin-holon eigenstates 
are very prevalent in the spectral decomposition of $\rho_{\!\expval{ij}}$,
the opposite trend between the associated eigenvalue and the overall behavior of
$\mathcal{N}_{\!\expval{ij}}^{\mathrm{F}t}$ shows that all the other hopping-induced
amplitudes contribute significantly to the negativity in the underdoped regime.
For this reason we do not believe that our analysis is strictly related to the Higgs
mechanism introduced in \cite{Sachdev_Pseudogap_Toporder} as an origin to pseudogap
physics. In our framework, the strongest entanglement signature associated to Mott
and pseudogap physics is the presence of short-ranged spin-singlet entanglement,
which is strictly killed at the metal-metal transition.

\newpage

\section{Summary and Outlook}

In this work, we resolved the two-site entanglement
of Hubbard systems in three terms, arising from physically 
distinct processes: an antiferromagnetic component, related
to the formation of spin singlets; a holon-doublon component,
describing the binding of spatially separated empty and fully
occupied sites; finally, a hopping-induced component, accounting
for all quantum amplitudes involving the hopping of single particles
from one site to the other.

For pure two-site states, such as the ground state of an isolated
Hubbard dimer, all three terms are rigorously associated to symmetry
labels in the fermionic partial transpose, directly descending from
the charge conservation law in the Hubbard Hamiltonian. 

For mixed two-site states, relevant for any finite temperature 
calculation and for pairs of sites embedded in larger systems, 
the symmetry labeling holds only under the parity superselection rule.
Hence, the antiferromagnetic and the holon-doublon terms constitute the
two accessible symmetry-resolved components of entanglement.
The hopping-induced entanglement can be evaluated beyond the symmetry
construction, as the portion of entanglement that is filtered out by
the superselection constraint. 

We derived concrete measures of these entanglement components both in
the framework of the relative entropy of entanglement, which gives a
highly accurate quantification of the two superselected terms, and
as entanglement negativities, providing an upper bound to the rate 
of distillation of spin-singlet, holon-doublon and hopping-induced
Bell pairs. The negativities generally overestimate the entanglement,
but preserve the qualitative behavior of the two superselected components.

We applied this analysis to zero-temperature, $2\times2$ CDMFT simulations
of the two-dimensional Hubbard model, in the paramagnetic regime.
At half-filling, where a Mott-Hubbard metal-insulator transition takes place upon increasing 
the interaction strength, we show that the hopping entanglement is the only
nonvanishing term in the metallic phase, 
while the Mott phase is dominated by an almost constant spin-singlet 
component. The holon-doublon entanglement has significant weight only
in the vicinity of the metal-insulator transition, reconnecting to early
proposals of a nearest-neighbor holon-doublon binding as a driving 
mechanism for Mott localization.

Doping the system, we find that the antiferromagnetic entanglement that 
dominates the Mott insulator directly extends into the pseudogap-like metallic
regime, and strictly vanishes at the first-order transition to the overdoped
Fermi liquid. The holon-doublon entanglement is also greatly suppressed in the
overdoped metal, but there is no strict vanishing up to the largest doping 
considered. The hopping-induced entanglement dominates the whole doping range,
except a small region close to half-filling. Its main pure-state component is
of the holon-spin type, but the evolution with doping can be explained only by
accounting for the cumulative effects of all the other hopping terms in the
two-site reduced density matrix.

In light of the clear-cut physical picture emerging from this symmetry-driven
analysis, we strongly advocate for investigation within complementary theoretical
and computational methods -- including accurate large-scale tensor networks, or neural quantum state variational approaches -- and also with analog quantum simulators based on ultra-cold
atomic gases. We note that a preliminary investigation, restricted to the
spin-singlet component of the two-site entanglement, has already been pushed
forward by some of us, for the high-temperature pseudogap regime of the two-dimensional Hubbard
model \cite{SSR_PG_ColdAtoms}. Remarkably, these experimental results demonstrate
that the antiferromagnetic entanglement, in contrast to classical antiferromagnetic
correlations, is confined to nearest neighbors and vanishes whenever the pseudogap
is closed, either by heating or overdoping. 
Extending the experimental, or numerically exact, analysis to the holon-doublon and
hopping entanglement will allow one to certify the robustness of our CDMFT results, at
least at the level of qualitative claims.

Finally, the investigation of superconducting regimes calls for a nontrivial 
extension of the analysis, to account for the generation of charge-nonconserving
terms in the reduced density matrix. 
We identify a viable strategy based on including all the anomalous amplitudes in a
pairing-driven entanglement term, similarly to what we have proposed in this work
for the hopping-induced entanglement, in mixed states. This approach would naturally
connect with the established framework of entanglement asymmetry, a genuine quantum
resource inherent to symmetry-broken phases
\cite{Vaccaro2008,Ares2023,Ares2024,Ares2025,Ares2026}.

\section*{Acknowledgments}
We acknowledge stimulating discussions with Marcello Dalmonte, 
regarding the intricate relationship between superselection rules
and different notions of symmetry resolution, and with Frederic
Bippus, Viktor Eisler and Alessio Paviglianiti, regarding the
significance of the fermionic partial transpose.
We are also grateful to Samuele Giuli and Edoardo Zavatti for 
support with the assessment of the weak-coupling limitations
of CDMFT with respect to CRISB.

\section*{Open Software and Data}
All numerical data discussed in the manuscript (CDMFT) and its appendix (CRISB), together with the scripts to compute entanglement measures and evaluate density matrices for the noninteracting limit are included in a publicly available open data repository \cite{data_repo}. The CDMFT/ED implementation, based on EDIpack
\cite{EDIpack,EDIpack2,EDIpack_code},
is also publicly available
\cite{code_repo}.

\clearpage

\appendix

\section{A note on the charge-imbalance blocks in the partial transpose}
\label{app:imbalance}

\begin{figure}
    \centering
    \includegraphics[width=.7\linewidth]{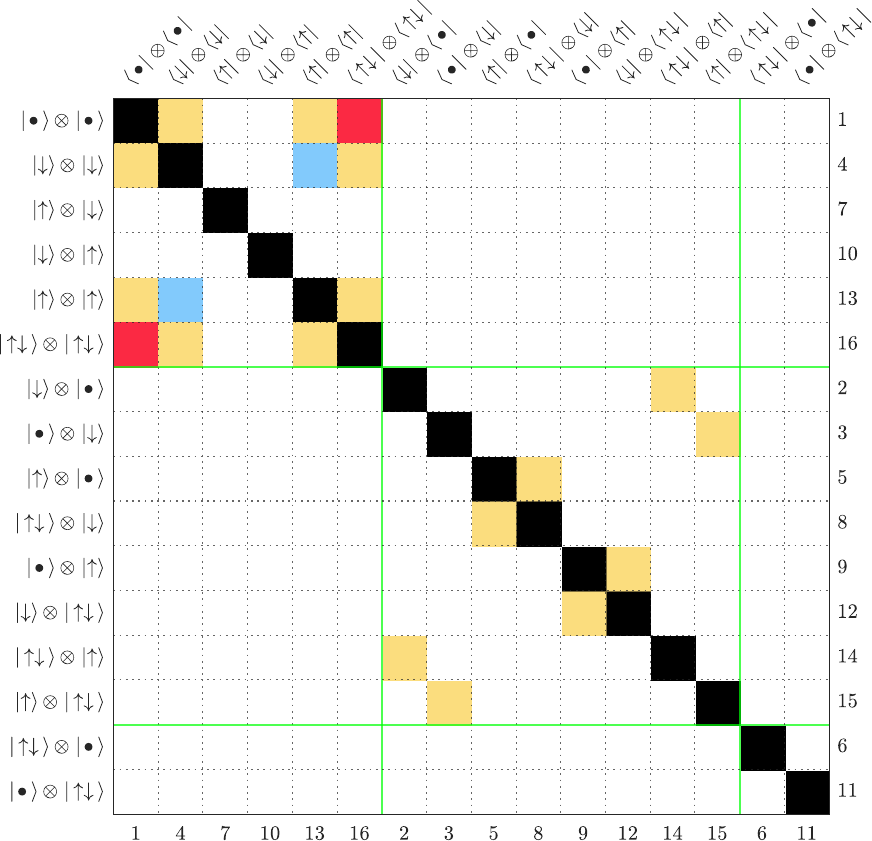}\\[1ex]
    \caption{Charge-imbalance sectors in the partial transpose of a generic
    mixed two-site density matrix, for a Hubbard system. In contrast to the
    symmetry resolution depicted in Fig.\,\ref{fig:mixed_rdm_and_ppt}, the
    charge-imbalance sectors are always well defined, even in absence of any
    superselection rule \cite{Imbalance_negativity_1st_paper,Murciano_imbalance_negativity_free_fermions,Murciano_imbalance_negativity_nonequilibrium}.
    Yet, hopping, antiferromagnetic and holon-doublon
    amplitudes are all mixed in the zero-imbalance sector, failing to discern
    the different physical origins of the two-site entanglement.}
    \label{fig:charge_imbalance}
\end{figure}

In Section \ref{sec:sym-res-mixed} we
have discussed the inherent difficulty in defining local charge-jump
sectors in the fermionic partial transpose of a mixed two-site
density matrix, in absence of any superselection rule. Hence, one 
could wonder if a more general notion of symmetry resolution
applies. Indeed, it is well known that in 
systems with conserved charges, the partial transpose of any bipartite density
matrix can be put in block diagonal form, with each symmetry sector 
labeled by the \emph{charge imbalance} between the two subsystems
\cite{Imbalance_negativity_1st_paper,Murciano_imbalance_negativity_free_fermions,Murciano_imbalance_negativity_nonequilibrium}.
For our two-site density matrix, such block-diagonal form is 
reported in Fig.\,\ref{fig:charge_imbalance}.
However, we immediately notice how the three physically distinct
sources of two-site entanglement are completely mixed in the 
zero-imbalance sector: it contains both the spin-singlet and 
holon-doublon contributions, plus the hopping amplitudes generated
by the mixing. This is not surprising, as those same four hopping 
amplitudes are the ones connecting the antiferromagnetic and the
holon-doublon sectors in Fig.\,\ref{fig:mixed_rdm_and_ppt}.
The remaining hopping amplitudes are contained in the 1-imbalance
sector, while the 2-imbalance sector is \emph{inactive} for it
has no off-diagonal elements.

Given that all the three different kinds of off-diagonal amplitudes
that we propose to probe different aspects of Hubbard physics are all
mixed in the same charge-imbalance sector, we argue that such a very 
general notion of symmetry resolution for the partial transpose is not
particularly insightful in this context. In fact, both for the 
entanglement negativity \cite{Murciano_imbalance_negativity_free_fermions}
and for the quantum relative entropy or other distance-based measures of
entanglement \cite{Capizzi_sym-resolved_QRE_CFT}, it is generally known
that all the active charge-imbalance sectors contribute equally to the
full entanglement, at leading order, so that any imbalance sector can be
used to infer the full entanglement in the system. While this 
\emph{equipartition} of entanglement in charge-imbalance sectors is 
indeed a remarkable property of charge-conserving Hamiltonians, it does
not give a genuinely differentiated access to complementary sources of
entanglement in the system. This is in turn the main goal of this work,
as we thoroughly demonstrate in the main text.

\section{Plaquette CDMFT/ED}
\label{app:cdmft_ed}

\begin{figure}
    \centering
    \includegraphics[width=0.66\linewidth]{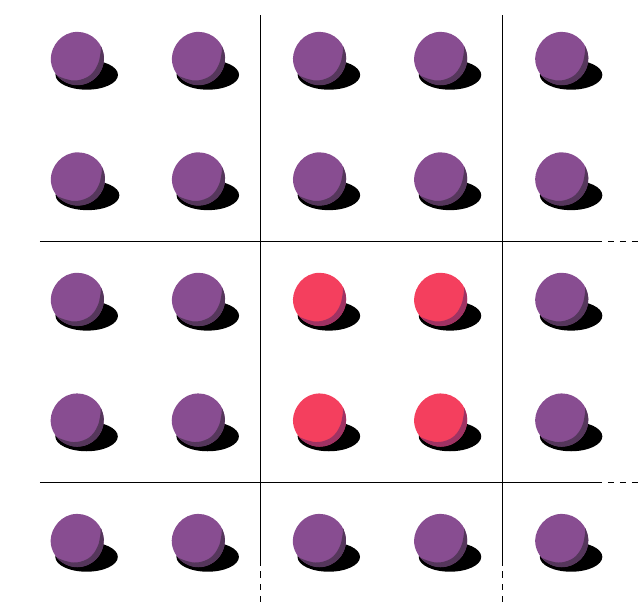}\\[5ex]
    \includegraphics[width=0.66\linewidth]{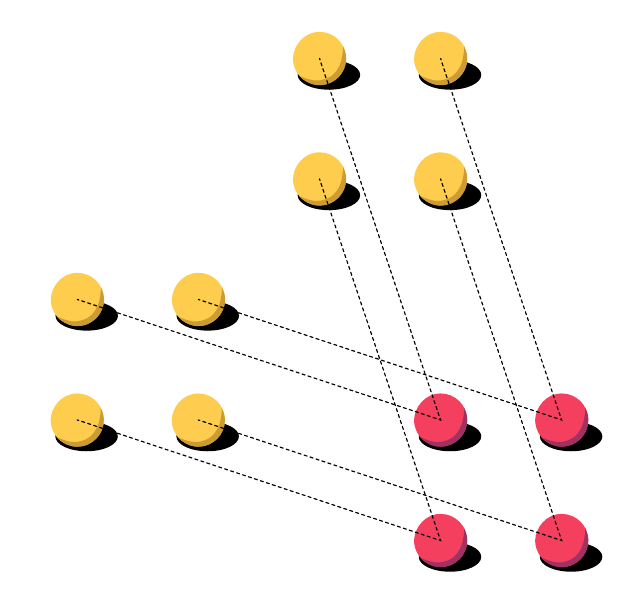}\\[2ex]
    \caption{Schematic of the CDMFT construction. In the
    top panel we show the tiling of the square lattice,
    identifying in red a single $2\times2$ plaquette.
    In the bottom panel we illustrate the auxiliary 
    impurity model, with an impurity cluster (in red)
    subjected to the same Hamiltonian as the original 
    lattice, and two noninteracting copies of it
    (\emph{replicas}, in yellow), subjected to 
    free-to-optimize kinetic terms $t_b$ and 
    $\mu_b$. The dashed lines represent diagonal
    hybridization amplitudes $V_{i,b}$, where $i$ spans
    the index of the impurity cluster and $b$ spans the
    replica index. See text for details.
    }
    \label{fig:cdmft_scheme}
\end{figure}

As represented in the top panel of Fig.\,\ref{fig:cdmft_scheme}, in cellular
dynamical mean-field theory one tiles the original, unbounded, lattice into
a periodic repetition of finite, bounded, clusters. These are embedded in
(\ie hybridized with) an effective noninteracting bath, to be determined 
self-consistently to satisfy the relationship 
\begin{equation}
    \mathbf{G}_\mathrm{cluster}(i\omega) = 
    \sum_{\mathbf{k}\in\mathrm{RBZ}} \mathbf{G}_\mathrm{lattice}(i\omega,\mathbf{k}),
\end{equation}
where RBZ indicates the reduced Brillouin zone associated to the periodic tiling
and the Green's function matrices span the degrees of freedom inside each cluster.
To work at zero temperature, the effective bath is best represented as a discrete
lattice Hamiltonian, so that the ``cluster+bath'' impurity model can be solved with 
exact diagonalization \cite{DMFT/ED_Caffarel,DMFT/ED_Capone}. In contrast to simpler
single-orbital/single-site schemes, the preservation of lattice symmetries demands
a careful parametrization of the bath Hamiltonian \cite{Koch2008PRB,Diego_CDMFTbath}. 
Among the different possible choices, we group the bath sites into \emph{noninteracting}
\emph{replicas} of the cluster. Formally
\begin{align}
     &H_\mathrm{ED} = H_\mathrm{cluster} + H_\mathrm{bath} + H_\mathrm{hybr.}\\[1ex]
&H_\mathrm{cluster} = -t \sum_{\expval{ij}}\sum_{\sigma} c^\dagger_{i,\sigma} c_{j,\sigma} 
+ \sum_i \left[U\left(c^\dagger_{i,\uparrow} c_{i,\uparrow} - \frac{1}{2}\right)
                 \left(c^\dagger_{i,\downarrow} c_{i,\downarrow} - \frac{1}{2}\right)
            -\mu \sum_\sigma c^\dagger_{i,\sigma}c_{i,\sigma}\right]\label{eq:H_cluster}\\[1ex]
&H_\mathrm{bath} = \sum_b \Biggl[
    -t_b \sum_{\expval{ij}}\sum_{\sigma} d^\dagger_{i,b,\sigma} d_{j,b,\sigma}
    -\mu_b \sum_{i,\sigma} d^\dagger_{i,b,\sigma}d_{i,b,\sigma}
    \Biggr] \\[1ex]
&H_\mathrm{hybr.} = \sum_i \sum_b \left[V_{i,b} \sum_\sigma 
                    \left( c^\dagger_{i,\sigma} d_{i,b,\sigma} +
                           d^\dagger_{i,b,\sigma} c_{i,\sigma}\right)
                                  \right].
\end{align}
Since these replicas have the same Hamiltonian as the original lattice -- except for 
the Hubbard interaction -- a diagonal and uniform hybridization with the original 
cluster $V_{i,b} = V_b$ will ensure a perfect preservation of the 
original lattice symmetries (see bottom panel in Fig.\,\ref{fig:cdmft_scheme}).
To accelerate the convergence of the self-consistent cycle, we allow the hybridization
amplitudes to be differentiated between different sites at each step. The final 
self-consistent values of these hybridization amplitudes have been checked to be 
uniform for each replica, within numerical accuracy, resulting in a symmetrical 
$\mathbf{G}_\mathrm{cluster}(i\omega)$ matrix and symmetrical two-site density
matrices: each of the four nearest-neighbor pairs is described by numerically 
indistinguishable density matrices $\rho_{\!\expval{ij}}$, and two numerically
equivalent $\rho_{\!\expval{\expval{ij}}}$ are associated to the two next-nearest
neighbor pairs. A script explicitly checking the equivalence of these density
matrices is included in the open data repository \cite{data_repo}.

\smallskip

The Lanczos diagonalization of $H_\mathrm{ED}$ is performed with a cluster
generalization of the massively-parallel algorithm introduced in 
\textsc{EDIpack} \cite{EDIpack},
exploiting the conservation of the total $N_\uparrow$ and $N_\downarrow$,
as guaranteed, in the normal phase, by the symmetries of the Hubbard model.
The algorithm employed to efficiently trace the bath degrees of freedom 
from the ground state of $H_\mathrm{ED}$ is thoroughly described in 
\cite{EDIpack2}. Further traces inside the cluster are finally performed
to get $\rho_{\!\expval{ij}}$ and $\rho_{\!\expval{\expval{ij}}}$.
All the relevant Fortran subroutines can be found in \cite{code_repo}.

\section{CDMFT vs CRISB in the weak-coupling limit}
\label{app:risb}

The attentive reader must have noticed that the exact values of
full and symmetry resolved measures of entanglement provided in
Section \ref{sec:dimer}, for the noninteracting infinite square
lattice, do not quantitatively match the $U\ll D$ limit of our 
CDMFT simulations. In principle, the DMFT approximation is
exact for a noninteracting system: the self-energy is
identically zero, so indeed local \cite{DMFT_RMP}.
However, once numerical solutions of the auxiliary impurity
problem are involved, additional considerations apply. In the
context of Hamiltonian based impurity solvers, like the numerically
exact Lanczos diagonalization we employ in this work 
\cite{EDIpack,EDIpack2,EDIpack_code,code_repo}, the control 
parameter for accuracy is the size of the bath. Given the exponential
scaling of exact diagonalization techniques, our implementation cannot
treat bath sizes much larger than the impurity, despite its
massive parallelization \cite{EDIpack,code_repo}. For a $2\times2$
plaquette, the restriction goes as tight as two or three copies of
the cluster. Furthermore, the replica parametrization (see Appendix
\ref{app:cdmft_ed}) enforces the number of bath sites to be even, which
disfavors the convergence of phases with a large spectral weight
at the Fermi level \cite{DMFT_RMP}. 
For all these reasons, we expect the weak-coupling regime to be the
least accurate, among all the phases we address in the paper.

\smallskip

\begin{figure}
    \centering
    \includegraphics[width=.85\linewidth]{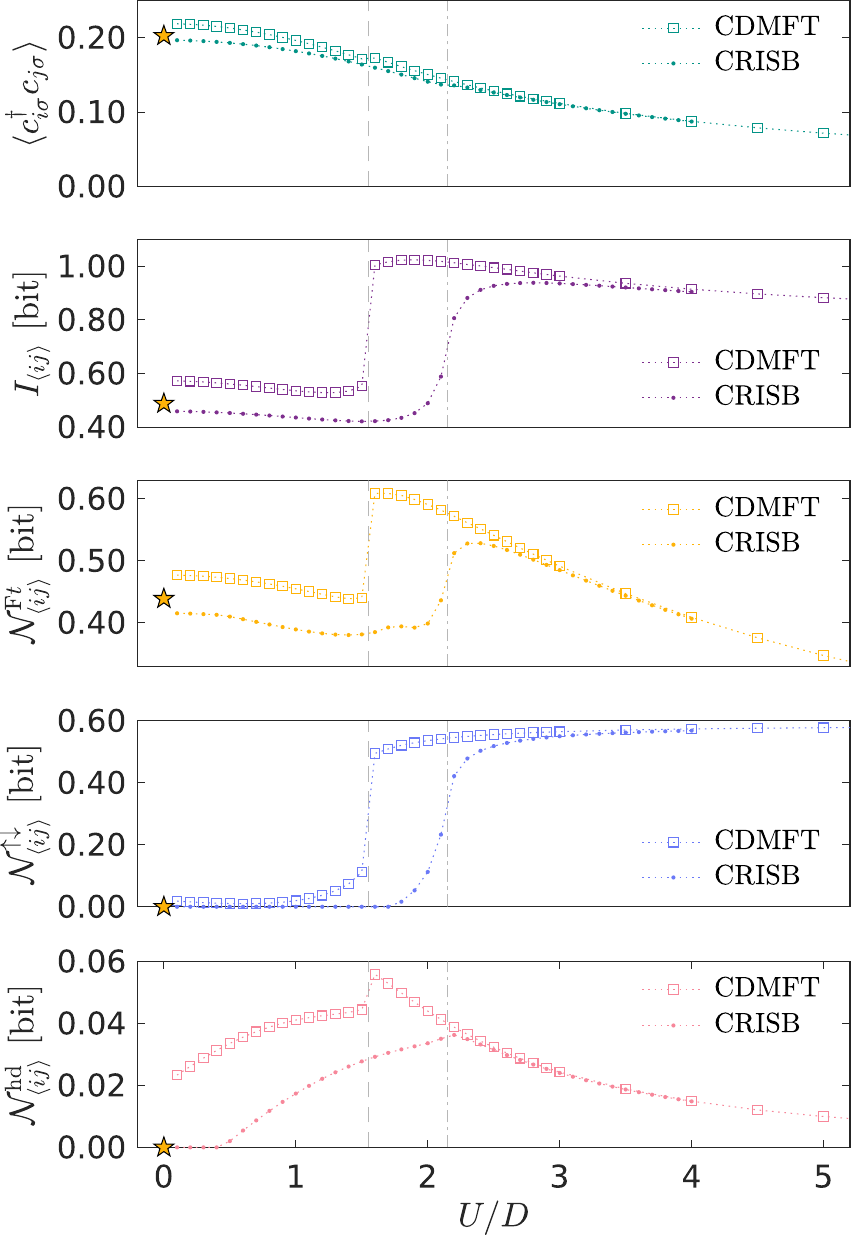}
    \caption{Comparison of various static quantities
    between CDMFT and CRISB, across the interaction-driven
    Mott transition, at half-filling. In the $U\ll D$ 
    limit, CRISB achieves a much better agreement with 
    the exact noninteracting values (stars), as the
    method directly optimizes for the one-body correlation
    matrix, from which all quantities are derived, in the
    free case. Dashed vertical lines indicate the
    Mott transition in CDMFT. Dash-dotted lines indicate
    the Mott transition in CRISB. When both calculations
    are in the Mott phase, an excellent agreement is 
    quickly recovered.}
    \label{fig:CRISBvCDMFT}
\end{figure}

To quantify the shortcomings of the finite CDMFT bath, beyond the
noninteracting limit, we compare in Fig.\,\ref{fig:CRISBvCDMFT} the
CDMFT/ED data from Section \ref{sec:half_filled_MIT} in the main text
with calculations obtained with an alternative cluster embedding theory:
the cluster rotationally-invariant slave boson approach (CRISB).
The method, introduced by Lechermann et al. \cite{RISBoriginal}, is
known to reproduce the Gutzwiller approximation in the limit
of a single-site cluster \cite{Gebhard_RISB=Gutzw}, but is able to
capture nontrivial multiorbital and/or nonlocal correlation effects
in many systems 
\cite{RISB_Isidori2009,GhostPlutonium,Mejuto_gGut_multiorbital,MejutoZaera2024_ghostMolecules,Tagliente2025,Giuli_Altermagnet_gRISB,TopGhost}, 
including the normal and $d$-wave phases of multi-layered cuprates 
\cite{Bi2223_CRISB}. In its quantum embedding formulation
\cite{GhostEmbedding}, the RISB approach has been shown to 
reproduce exactly the DMFT self-consistency if the bath 
size is sent to infinity \cite{GhostVsDMFT,Giuli2026_gRISB=DMFT}.

More importantly for the present discussion, \emph{in the limit of small
bath sizes}, the RISB self-consistency ensures a much easier convergence
of \emph{static} quantities, which determine all entanglement features
in the system (entanglement is an equal-time property 
\cite{EntanglementRMP,Laflorencie_Review,Calabrese_EEreview}).
On the other hand, DMFT is based on a \emph{dynamical} self-consistency,
so that static quantities are pinned only through a summation over
Matsubara frequencies \cite{DMFT_RMP}, which is affected by long tail
effects and issues in reproducing the low-frequency behavior,
when the bath is truncated \cite{Capone2004_1dCDMFT}. 

In fact, the nearest-neighbor one-body correlation matrix 
$\gamma_{\!\expval{ij}}$ -- which in the noninteracting limit
(Eq.\,\ref{eq:free_gamma}) fully determines the nearest-neighbor
density matrix $\rho_{\!\expval{ij}}$ -- is a fully self-consistent
object in the CRISB formalism, while it is determined by a frequency 
sum of the self-consistent cluster Green's function in CDMFT. 

Concretely, in the top panel of 
Fig.\,\ref{fig:CRISBvCDMFT}, we show the off-diagonal component of 
$\gamma_{\!\expval{ij}}$, as a function of $U/D$ (the diagonal is 
trivially constant).
It is evident how the exact noninteracting limit 
(Eq.\,\ref{eq:free_gamma}, and star symbol in 
Fig.\,\ref{fig:CRISBvCDMFT}) is much better approximated 
by CRISB in the $U\ll D$ calculations, with respect to CDMFT.
The residual numerical error in CRISB can be attributed to the
discretization of the momentum grid, corresponding to a mild
finite size effect: $100\times100$ momenta, versus an infinite 
square lattice. The same grid in CDMFT leads to a self-consistent
cluster Green's function with a clearly worse static component.

\smallskip

In the weak-coupling
regime, we can expect the Gaussian approximation of the system to be
reliable \cite{bellomia_local_classical}, so that the evolution of
$\gamma_{\!\expval{ij}}$ gives a fair indication on the evolution of
$\rho_{\!\expval{ij}}$. Once again, both methods converge to a smooth
evolution of the nearest-neighbor correlation function, signaling a 
well-behaved self-consistency loop. Yet, CDMFT significantly  
overestimates this quantity with respect to CRISB. As a consequence,
all the measures of full and symmetry-resolved two-site entanglement
are overestimated in CDMFT, as shown in the other panels of 
Fig.\,\ref{fig:CRISBvCDMFT}. Remarkably, while the mutual information 
and the hopping negativity display the same behavior, with just a
nearly-rigid quantitative shift, the two components of the superselected
entanglement feature a striking qualitative disagreement. Both the
spin-singlet and the holon-doublon negativities are strictly nonzero in
CDMFT, even for very small interaction strengths, signaling a non-positive
partial transpose. On the contrary, in CRISB the PPT criterion successfully
detects the vanishing of superselected entanglement in the noninteracting
limit and suggests that this vanishing persists also at low to moderate $U/D$ ratios.
We recall that, given the dimension of the spin-singlet and holon-doublon
sectors in the partial transpose, the PPT criterion is exact in both 
directions for the superselected entanglement, so the vanishing of these
two negativities implies also the exact vanishing of the more quantitative
REE measures. Comparing with Fig.\,\ref{fig:half_filling}, we see how the 
failure of the relative entropy in witnessing the weak-coupling death of
superselected entanglement in CDMFT is much less noticeable, but we remark
that the values, albeit very small ($\lesssim 10^{-3}$ for sufficiently 
small interaction values, see also \cite{Bellomia2023}), are definitely
above the numerical noise and so to be regarded as finite. 

\smallskip

After the Mott transition, which happens at different interaction strengths
in CDMFT and CRISB (see vertical dashed lines in Fig.\,\ref{fig:CRISBvCDMFT})
all quantities quickly converge to an excellent agreement between the CRISB
and the CDMFT calculations, reconciling with the notion that an even number
of bath sites is much better suited to simulate gapped phases at strong
coupling, even when complex many-body static quantities are concerned.
We expect the CDMFT simulation in the doping-driven MIT (Section \ref{sec:doped_MIT})
to also be faithful, as it reproduces many features of established
low-temperature calculations with an infinite bath, including effects 
on static quantities: compare the first-order jumps in the single-site 
density and entropy in \cite{BellomiaPhD} and \cite{WalshPRXQuantum2020} 
(see also \cite{Sordi_njumps_PRL,Sordi_njumps_PRB,Gingras_njumps_DCA}).


\bibliography{main.bib}

\end{document}